\documentclass[5p,10pt]{elsarticle}

\usepackage{amsmath}
\usepackage{amssymb}
\usepackage{amsfonts}
\usepackage{exscale}
\usepackage{graphicx}
\usepackage{booktabs}
\usepackage[tableposition=top,font=footnotesize,labelfont=bf]{caption}
\usepackage{subfig}
\usepackage{dsfont}
\usepackage{xcolor}
\usepackage{soul}

\newcommand{\R}{{\mathbb{R}}}

\newtheorem{theorem}{Theorem}[section]

\newtheorem{definition}[theorem]{Definition}

\newtheorem{remark}[theorem]{Remark}
\newtheorem{assumption}[theorem]{Assumption}
\numberwithin{equation}{section}

\def\L2{{\cal L}_2}
\def\L2e{{\cal L}_{2e}}

\def\rea{\mathbb{R}}

\def\diag{\mbox{diag}}

\def\col{\mbox{col}}

\def\begequarr{\begin{eqnarray}}
\def\endequarr{\end{eqnarray}}
\def\begequarrs{\begin{eqnarray*}}
\def\endequarrs{\end{eqnarray*}}
\def\begarr{\begin{array}}
\def\endarr{\end{array}}
\def\begequ{\begin{equation}}
\def\endequ{\end{equation}}
\def\begs{\begin{split}}
\def\lab{\label}
\def\begdes{\begin{description}}
\def\enddes{\end{description}}
\def\begenu{\begin{enumerate}}
\def\begite{\begin{itemize}}
\def\endite{\end{itemize}}
\def\endenu{\end{enumerate}}

\def\lef[{\left[\begin{array}}
\def\rig]{\end{array}\right]}

\def\begcen{\begin{center}}
\def\endcen{\end{center}}
\def\begrem{\begin{remark}\rm}
\def\endrem{\end{remark}}

\begin{document}
\begin{frontmatter}
\title{\LARGE \textbf{A survey on modeling of microgrids---from fundamental physics to phasors and voltage sources}}
%
\author[sch]{Johannes Schiffer\corref{cor1}}\ead{j.schiffer@leeds.ac.uk}  
\author[zon]{Daniele Zonetti}\ead{zonetti@lss.supelec.fr}               
\author[zon]{Romeo Ortega}\ead{ortega@lss.supelec.fr}               
\author[stan]{Aleksandar Stankovi\'{c}}\ead{astankov@ece.tufts.edu}  
\author[sez]{Tevfik Sezi}\ead{tevfik.sezi@arcor.de}  
\author[rai]{J\"org Raisch}\ead{raisch@control.tu-berlin.de}  
		\cortext[cor1]{Corresponding author J.~Schiffer. Tel. +44 (0)113 343 9719.
			Fax +44 (0)113 343 2032.}
		\address[sch]{School of Electronic and Electrical Engineering, University of Leeds, Leeds LS2 9JT, UK}  
\address[zon]{Laboratoire des Signaux et Syst\'{e}mes, \'{E}cole Sup\'{e}rieure d$'$Electricit\'{e} (SUPELEC), Gif-sur-Yvette 91192, France}             
\address[stan]{Tufts University, Medford, MA 02155, USA}
\address[sez]{Siemens AG, Smart Grid Division, Energy Automation, Humboldtstr. 59, 90459 Nuremberg, Germany}  
\address[rai]{Technische Universit\"at Berlin, Einsteinufer 11, 10587 Berlin, Germany and Max-Planck-Institut f\"ur Dynamik komplexer technischer Systeme, Sandtorstr. 1, 39106 Magdeburg, Germany}
%


\begin{abstract}
Microgrids have been identified as key components of modern electrical systems to facilitate the integration of renewable distributed generation units.
Their analysis and controller design requires the development of advanced (typically model-based) techniques naturally posing an interesting challenge to the control community. Although there are widely accepted reduced order models to describe the dynamic behavior of microgrids, they are typically presented without details about the reduction procedure---hampering the understanding of the physical phenomena behind them. Preceded by an introduction to basic notions and definitions in power systems, the present survey reviews key characteristics and main components of a microgrid. We introduce the reader to the basic functionality of DC/AC inverters, as well as to standard operating modes and control schemes of inverter-interfaced power sources in microgrid applications. Based on this exposition and starting from fundamental physics, we present detailed dynamical models of the main microgrid components. Furthermore, we clearly state the underlying assumptions which lead to the standard reduced model with inverters represented by controllable voltage sources, as well as static network and load representations, hence, providing a complete modular model derivation of a three-phase inverter-based microgrid. 
\end{abstract}
%
\begin{keyword}
Microgrid modeling \sep microgrid analysis \sep smart grid applications \sep inverters 
\end{keyword}
\end{frontmatter}
%
%
\section{Introduction}
%
\subsection{Motivation}
%
It is a widely accepted fact that fossil-fueled thermal power generation highly contributes to greenhouse gas emissions \cite{lund07,machowski08,lund09}.
In addition, a growing stream of scientific results \cite{houghton96,hansen05,solomon07} has substantiated claims that these emissions are a key driver for climate change and global warming. As a consequence, many countries have agreed to reduce their greenhouse gas emissions. 

Apart from a reduction of energy consumption, {\em e.g.}, through an increase in efficiency,
one possibility to reduce greenhouse gas emissions is to shift the energy production from fossil-fueled plants towards renewable sources \cite{lund07,lund09,chowdhury09}.
Therefore, the worldwide use of renewable energies has increased significantly in recent years \cite{teodorescu11}. 

Unlike fossil-fueled thermal power plants, the majority of renewable power plants are relatively small in terms of their generation power. 
An important consequence of this smaller size is that most of them are connected to the low voltage (LV) and medium voltage (MV) levels. Such generation units are commonly denoted as distributed generation (DG) units \cite{ackermann01}. In addition, most renewable DG units are interfaced to the network via DC/AC inverters. 
The physical characteristics of such power electronic devices largely differ from the characteristics of synchronous generators (SGs), which are the standard generating units in existing power systems. Hence, different control and operation strategies are needed in networks with a large amount of renewable DG units~\cite{green07,varaiya11, teodorescu11}. 
\subsection{The microgrid concept}
%
One potential solution to facilitate the integration of large shares of renewable DG units are microgrids \cite{lasseter02,hatziargyriou07,green07,katiraei08,chowdhury09,strbac15}. A microgrid gathers a combination of generation units, loads and energy storage elements at distribution or sub-transmission level into a locally controllable system, which can be operated  either in grid-connected mode or in islanded mode, {\em i.e.}, in a completely isolated manner from the main transmission system. The microgrid concept has been identified as a key component in future electrical networks~\cite{chowdhury09,path_microgrid,lasseter11,strbac15}. 

Many new control problems arise for this type of networks. Their satisfactory solution requires the development of advanced model-based controller design techniques that often go beyond the classical linearization-based nested-loop proportional-integral (PI) schemes. This situation has, naturally, attracted the attention of the control community as it is confronted with some new challenging control problems of great practical interest. 

It is clear that to carry out this task it is necessary to develop a procedure for assembling mathematical models of a microgrid that reliably capture the fundamental aspects of the problem. 
Such models have been developed by the power systems and electronics communities and their pertinence has been widely validated in simulations and applications \cite{coelho02,katiraei06,pogaku07,mohamed08}. However, these are reduced or simplified, {\em i.e.}, linearized, models that are typically presented without any reference to the reduction procedure---hampering the understanding of the physical phenomena behind them.

\subsection{Existing literature}
%
For the purposes of this survey, previous work on microgrid modeling can be broadly categorized into two classes. The first class focusses on modeling and control of inverter-interfaced DG units in microgrid applications, but the model derivation is restricted to individual DG units and the current and power flows between different units are not considered explicitly \cite{green07,pogaku07,katiraei06,zhong12,rocabert12,guerrero13,bidram13,bidram14}. 
The second class discusses models of microgrids including electrical network interactions, but the model derivation is based on linearization ({\em i.e.}, the so-called small-signal model) \cite{katiraei07,pogaku07,mohamed08}.
Furthermore, this class of modeling is often tied to specific network control schemes, such as droop control \cite{coelho02,pogaku07} or to specific test networks \cite{katiraei06,katiraei07,mohamed08,miao11}. Building on this previous work and in a survey-like manner, the present paper brings both aforementioned classes together to formulate a generic modular model of a microgrid.  

Going beyond a mere review of existing microgrid models, we employ model reduction via a time-scale separation together with the subsequent derivation of the well-known power flow equations, which is a standard procedure in SG-based networks \cite{venkatasubramanian95}. A similar approach has also been employed for microgrids in \cite{mariani14,mariani14_2,riverso14,luo14}. However, neither reference provides a detailed model derivation for inverter-interfaced units. Also, the analysis in \cite{mariani14,mariani14_2} is restricted to an AC microgrid consisting of two inverters connected via a resistive-inductive line and two local resistive-inductive loads, while the modeling procedure of the present paper applies to networks with generic topo\-logy and arbitrary number of units. 
\subsection{About the survey}
%
The present survey is an attempt to provide a guideline for control engineers attracted by this fundamental application for Smart Grids to assess the importance of the main dynamical components of a three-phase inverter-based microgrid as well as the validity of different models used in the power literature.
To this end, we at first review some fundamental concepts and definitions in power systems, including a survey on the notion of instantaneous power. Subsequently, we introduce the reader to the microgrid concept and discuss its main components. We illustrate that inverter-interfaced units are the main new elements in future power networks, detail the basic functionality of inverters and review the most common operation modes of inverter-interfaced units together with their corresponding control schemes. This paves the path for---starting from fundamental physics---presenting detailed dynamical models of the individual microgrid components. Subsequently, we clearly state the underlying assumptions which lead to the standard reduced model with inverters represented by controllable voltage sources, as well as static network and load representations. This reduced model is used in most of the available work on microgrid control design and analysis \cite{simpson13,schiffer13_2,ainsworth13,muenz14,doerfler14}.

We focus on purely inverter-based networks, since in\-ver\-ter-interfaced units are the main new elements in microgrids compared to traditional power systems. However, we remark that the employed modeling and model reduction techniques can equivalently be applied to standard bulk power system models as well as to power systems with mixed generation pool. For modeling of traditional electro-mechanical SG-based units, the reader is referred to standard textbooks on power systems \cite{kundur94,machowski08,anderson02}.

The main contributions of the present survey paper are summarized as follows.
\begite
\item Provide a detailed comprehensive model derivation of a microgrid based on fundamental physics and combined with detailed reviews of the microgrid concept, its components and their main operation modes.
\item Answer the question, when an inverter can be modeled as a controllable AC voltage source and depict the necessary underlying model assumptions.
\item Show that the usual power flow equations can be obtained from a network with dynamic line models via a suitable coordinate transformation (cal\-led $dq$-transformation) together with a singular per\-tur\-ba\-tion argument.
\item By combining the two latter contributions, recover the reduced-order microgrid model currently widely used in the literature.
\endite
We emphasize that the aim of the present survey is not to give an overarching justification for the final (simplified) model, but to provide a comprehensive overview of the modeling procedure for main microgrid components together with their dynamics, as well as of the main necessary assumptions, which allow the reduction of model complexity.
Which of the presented models (if any) is appropriate for a specific control design and analysis cannot be established in general, but has to be decided by the user. Any model used in simulation and analysis necessarily involves certain assumptions. 
Therefore, it is of great importance that the user is aware of the pertinence of the employed model to appropriately assess the implications of a model-based analysis. 

The remainder of this survey is structured as follows. Basic preliminaries, such as definitions and transformations in power systems are given in Section~\ref{sec2}. The microgrid concept 
is reviewed in Section~\ref{sec:mg}. 
A detailed dynamical model of a microgrid is derived in Section~\ref{sec4}. In particular, common operation modes of inverter-interfaced units are discussed therein. The model reduction yielding models of inverters as AC voltage sources and phasorial power flow equations is conducted in Section~\ref{sec5}. The survey paper is wrapped-up with some conclusions and a discussion of topics of future research in Section~\ref{sec7}.\\

\noindent {\bf Notation.} We define the sets $\R_{\geq0}:=\{x \in \R|x\geq0 \}$, $\R_{>0}:=\{x \in \R|x>0 \}$ and $\mathbb S:=[0,2\pi).$ For a set $\mathcal V$, let $|\mathcal V|$ denote its cardinality and $[\mathcal V]^k$ denote the set of all subsets of $\mathcal V$ that contain $k$ elements.
For a set of, possibly unordered, positive natural numbers $\mathcal V=\{l,k,\ldots,n\},$ the short-hand $i\sim\mathcal V$ denotes $i=l,k,\ldots,n.$
Given a positive integer $n,$ we use $\underline 0_n$ to denote the vector of all zeros, $\mathds{1}_n$ the
vector with all ones and $\mathbf{I}_n$ the $n \times n$ identity matrix.
Let $x=\col(x_1,\dots,x_n)\in\mathbb C^n$ denote a column
vector with entries $x_i \in \mathbb C.$ Whenever clear from the context, we simply write $x=\col(x_i)\in\mathbb C^n$. The 2-norm of a vector $x\in\mathbb C^n$ is denoted by $\|x\|_2=\sqrt{|x_1|^2+\ldots+|x_n|^2}.$ Let $\diag(a_i)\in\mathbb C^{n\times n}$ denote a diagonal matrix with entries $a_i \in \mathbb C.$ 
Let $j$ denote the imaginary unit. The conjugate transpose of a matrix $A\in\mathbb C^{n\times n}$ is denoted by $A^*.$ For a function $f:\rea^n \to \rea$, $\nabla f$ denotes the transpose of its gradient. The operator $\otimes$ denotes the Kronecker product.

\section{Preliminaries and basic definitions}
\lab{sec2}
\subsection{Symmetric AC three-phase signals}
\lab{ss:ac}

\begin{definition}\cite{dinAC}
A signal $x:\R_{\geq 0}\to\R$ is said to be an AC signal if it satisfies the following conditions
\begenu
\item it is periodic with period $T\in\R_{>0},$ {\em i.e.},
\begequ
x(t)=x(t+nT),\quad \forall n\in\mathbb N,\quad \forall t\geq 0,\notag
\endequ
\item its arithmetic mean is zero, {\em i.e.},
\begequ
\int_{t}^{t+T} x(\tau)d\tau=0\quad \forall t\geq 0.\notag
\endequ
\endenu
\end{definition}

\begin{definition}
A signal $x:\R_{\geq 0}\to\R^3$ is said to be a three-phase AC signal if it is of the form
$$x_{ABC}=\col(x_A, x_B,x_C),$$
where $x_A:\R_{\geq 0}\to\R,$ $x_B:\R_{\geq 0}\to\R$ and $x_C:\R_{\geq 0}\to\R$ are AC signals.
\end{definition}

A special kind of three-phase AC signals are symmetric AC three-phase signals, defined below.

\begin{definition}\citep[Chapter 2]{akagi07}\label{3p}
A three-phase AC signal $x_{abc}:\R_{\geq0}\to\R^3$ is said to be symmetric if it can be described by 
\begequ
x_{abc}(t)=\begin{bmatrix}
            x_a(t)\\x_b(t)\\x_c(t)
           \end{bmatrix}=
A(t) \begin{bmatrix} \sin(\delta(t))\\\sin(\delta(t)-\frac{2\pi}{3})\\\sin(\delta(t)+\frac{2\pi}{3}) \end{bmatrix},
\notag
\endequ 
where $A:\R_{\geq 0}\to \R_{\geq 0}$ is called the amplitude and $\delta:\R_{\geq 0}\to \mathbb S$ is called the phase angle of the signal. 
\lab{def:sym}
\end{definition}

Clearly, from the preceding definition, a symmetric three-phase AC signal $x_{abc}$ can be described completely by two signals: its angle $\delta$ and its amplitude\footnote{To simplify notation the time argument of all signals is omitted, whenever clear from the context. The same applies to the definition of signals, {\em i.e.}, a signal $x:\R_{\geq0}\to\R,$ is defined equally as $x\in\R,$ whenever clear from the context.} $A.$

\begin{definition}\citep[Chapter 2]{akagi07}
A three-phase AC signal is said to be asymmetric if it is not symmetric. 
\end{definition}

\begin{definition}\citep[Chapter 3]{heuck13}
A three-phase AC electrical system is said to be symmetrically configured if a symmetrical feeding voltage yields a symmetrical current and vice versa.
\end{definition}

\begin{definition}\citep[Chapter 3]{heuck13}
A three-phase AC power system is said to be operated under symmetric conditions if it is symmetrically configured and symmetrically fed.
\end{definition}	

\begin{figure}
\centering
\subfloat[Symmetric three-phase AC signal with constant amplitude]{
\includegraphics[width=.45\linewidth]{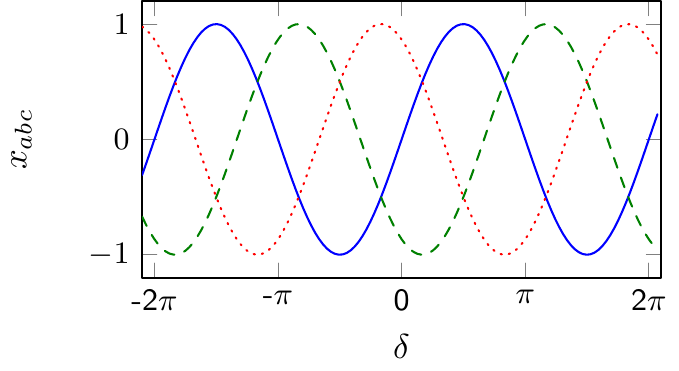}
\label{fig:symSig}}
\hfill
\subfloat[Symmetric three-phase AC signal with time-varying amplitude]{
\includegraphics[width=.45\linewidth]{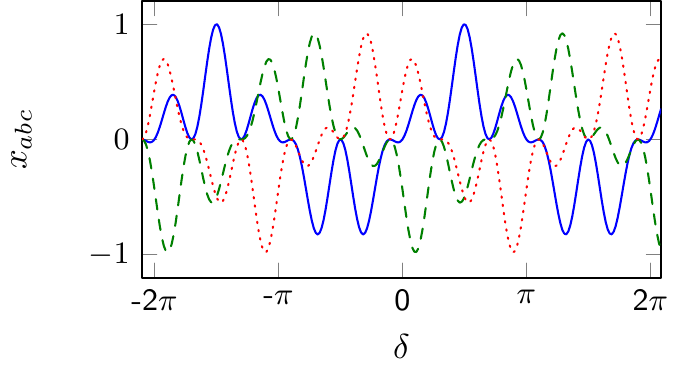}
\label{fig:symSigVar}}
\\
\subfloat[Asymmetric three-phase AC signal with phases not shifted equally by $\frac{2\pi}{3}$]{
\includegraphics[width=.45\linewidth]{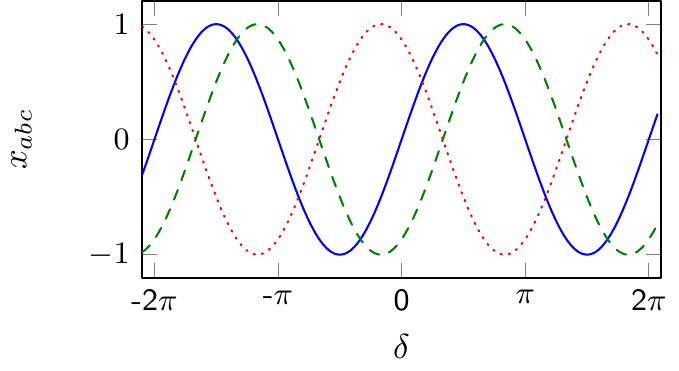}
\label{fig:unsymSig}}
\hfill
\subfloat[Asymmetric three-phase AC signal resulting of an asymmetric superposition of a symmetric signal with signals oscillating at higher frequencies]
{
\includegraphics[width=.45\linewidth]{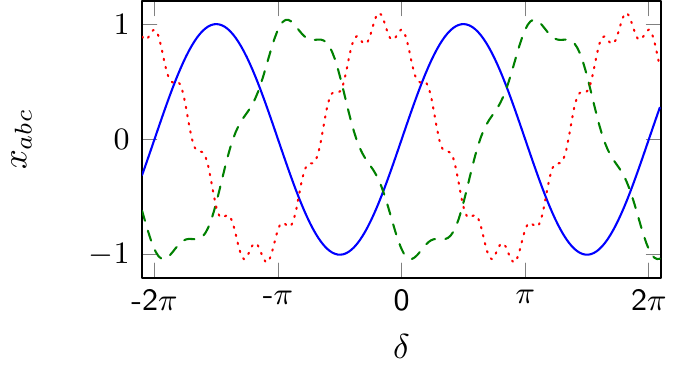}
\label{fig:distSig}}
\hfill
\caption[Symmetric and asymmetric AC three-phase signals]{Symmetric and asymmetric AC three-phase signals. The lines correspond to $x_a$ '\textcolor{blue}{---}', $x_b$ '\textcolor{black!50!green}{- -}', $x_c$ '\textcolor{red}{$\cdots$}'.}
\lab{fig:symSigs}
\end{figure}

Examples of symmetric and asymmetric three-phase AC signals\footnote{In this paper only AC systems and signals are considered. Therefore, the qualifier AC is dropped from now on.} are given in Fig.~\ref{fig:symSigs}. Note that the signal in Fig.~\ref{fig:symSigVar} satisfies Definition~\ref{3p} and only differs from the signal in Fig.~\ref{fig:symSig} in that it possesses a time-varying periodic amplitude.

\begrem
The terms ``balanced'' and ``unbalanced'' are frequently used as synonyms of ``symmetric'', respectively ``asymmetric'' in the literature \cite{akagi07,glover11}.
\endrem

\begrem
Three-phase electrical power systems consist of three main conductors in parallel. Each of these conductors carries an AC current. 
A three-phase system can be arranged in $\Delta$- or Y-configuration, see Fig.~\ref{fig:wyedelta}. The latter is also called wye-configuration.
Frequently, in a system with Y-configuration an additional fourth (grounded) neutral conductor is used to reduce transient overvoltages and to carry asymmetric currents \citep[Chapter 2]{glover11}. Such systems are typically called three-phase four-wire systems. Most three-phase power systems are four-wire Y-connected systems with grounded neutral conductor \cite[Chapter 2]{glover11}. However, it can be shown that, under symmetric operating conditions, this fourth wire does not carry any current and can therefore be neglected \citep[Chapter 2]{glover11}. 
\endrem

\begin{figure}
\includegraphics[width=1\linewidth]{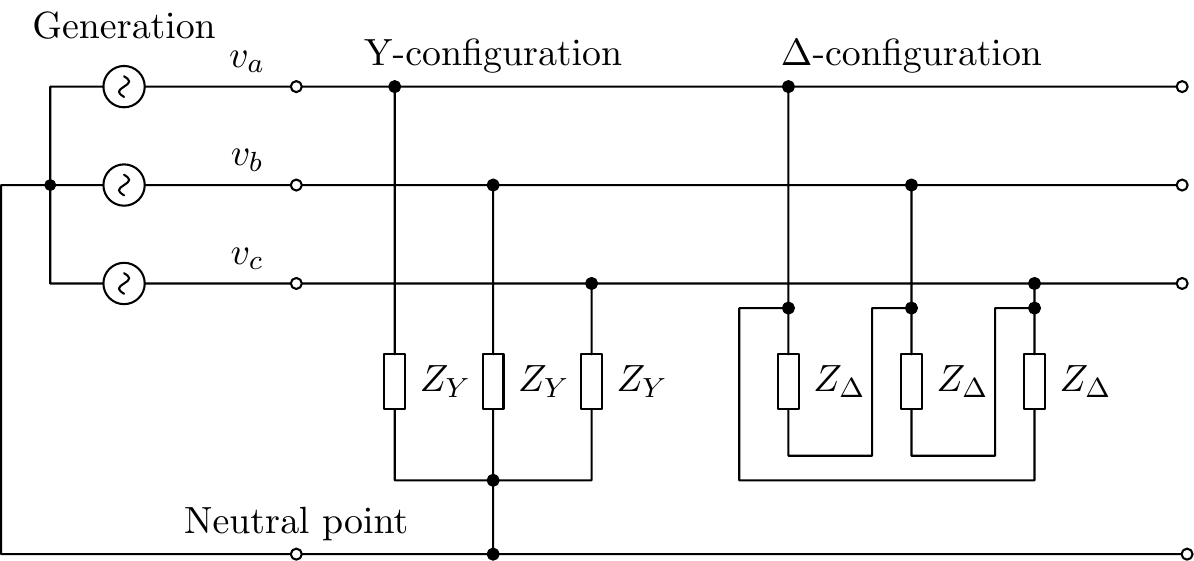}
\caption{Standard Y- and $\Delta$-configurations of three-phase AC power systems based
on \citep[Chapter 3]{heuck13}. $Z_\text{Y}\in\mathbb C$ denotes the impedance in Y-configuration, while $Z_\Delta\in\mathbb C$ denotes the impedance in $\Delta$-configuration.} 
\lab{fig:wyedelta}
\end{figure}

\subsection{The $dq0$-transformation}
\lab{ss:dq}
An important coordinate transformation known as $dq0$-transformation in the literature \cite{park29,paap00,anderson02,machowski08,teodorescu11,andersson12,zhong12} is introduced. 

\begin{definition}\citep[Chapter 4]{anderson02}, \citep[Chapter 11]{machowski08}
Let $x:\R_{\geq0}\to\R^3$ and $\varrho:\R_{\geq0}\to \mathbb S.$ Consider the mapping $T_{dq0}: \mathbb S\rightarrow \R^{3\times 3},$
\begequ
T_{dq0}(\varrho):=\sqrt{\frac{2}{3}}\begin{bmatrix}\cos(\varrho) & \cos(\varrho-\frac{2}{3}\pi) & \cos(\varrho+\frac{2}{3}\pi) \\
\sin(\varrho) & \sin(\varrho(t)-\frac{2}{3}\pi) & \sin(\varrho+\frac{2}{3}\pi)\\
\frac{\sqrt{2}}{2} & \frac{\sqrt{2}}{2} & \frac{\sqrt{2}}{2}
\end{bmatrix}.
\lab{tdq0}
\endequ
Then, $f_{dq0}:\R^3\times\mathbb S\to\R^3,$
\begequ
f_{dq0}(x(t),\varrho(t))=T_{dq0}(\varrho(t))x(t)
\lab{tdq2}
\endequ
is called $dq0$-transformation.
\end{definition}
Note that the mapping \eqref{tdq0} is unitary, {\em i.e.}, $T_{dq0}^\top = T_{dq0}^{-1}.$ From a geometrical point of view, the $dq0$-transformation is a concatenation of two rotational transformations, see \cite{paap00} for further details.
The variables in the transformed coordinates are often denoted by $dq0$-variables. 

The $dq0$-transformation offers various advantages when analyzing and working with power systems and is therefore widely used in applications \cite{paap00,anderson02,akagi07,teodorescu11,zhong12}. 
For example, the $dq0$-transformation permits, through appropriate choice of $\varrho$, to map three-phase AC signals to constant signals, {\em i.e.}, to transform periodic orbits into constant equilibria. This simplifies the control design and analysis in power systems, which is the main reason why the transformation \eqref{tdq2} is introduced in the present case. 
In addition, the transformation \eqref{tdq2} exploits the fact that, in a power system operated under symmetric conditions, a three-phase signal can be represented by two quantities. To see this, let $x_{abc}:\R_{\geq0}\to \R^3$ be a symmetric three-phase signal with amplitude $A:\R_{\geq0}\to \R_{\geq 0}$ and phase angle $\theta:\R_{\geq0}\to \mathbb S$, as in Definition \ref{3p}. Applying the mapping \eqref{tdq0} with some angle $\varrho:\R_{\geq0}\to\mathbb S$ to $x_{abc}$ yields
\begequ
x_{dq0}=\begin{bmatrix}
          x_d\\x_q\\x_0
         \end{bmatrix}=
T_{dq0}(\varrho)x_{abc}=\sqrt{\frac{3}{2}}A \begin{bmatrix} \sin(\theta-\varrho)\\\cos(\theta-\varrho)\\0
                         \end{bmatrix}.\notag
\endequ
Hence, $x_0=0$ for all $t\geq0.$ Therefore and as in this work only symmetric three-phase signals are considered, it is convenient to introduce the mapping
$T_{dq}: \mathbb S\rightarrow \R^{2\times 3},$
\begequ
T_{dq}(\varrho):=\sqrt{\frac{2}{3}}\begin{bmatrix}\cos(\varrho) & \cos(\varrho-\frac{2}{3}\pi) & \cos(\varrho+\frac{2}{3}\pi) \\
\sin(\varrho) & \sin(\varrho-\frac{2}{3}\pi) & \sin(\varrho+\frac{2}{3}\pi)
\end{bmatrix},
\lab{tdq}
\endequ
which, when applied to the symmetric three-phase signal $x_{abc}$ defined above, yields
\begequ
x_{dq}=\begin{bmatrix}
          x_d\\x_q
         \end{bmatrix}=
T_{dq}(\varrho)x_{abc}=\sqrt{\frac{3}{2}}A \begin{bmatrix}\sin(\theta-\varrho)\\\cos(\theta-\varrho)
                         \end{bmatrix}.\notag
\endequ
In the following, $x_{dq}$ are referred to as the $dq$-coordinates of $x_{abc}.$ Note that $x_{abc}=T_{dq}(\varrho)^\top x_{dq}.$

\begrem
There are several variants of the mapping \eqref{tdq0} available in the literature. They may differ from the mapping \eqref{tdq0} in the order of the rows and the sign of the entries in the second row of the matrix given in \eqref{tdq0}, see \cite{anderson02,andersson12,zhong12}. 
However, all representations are equivalent in the sense that they can all be represented by $T_{dq0}$ as given in \eqref{tdq0} by choosing an appropriate angle $\varrho$ and, possibly, rearranging the row order of the matrix $T_{dq0}.$ The same applies to the mapping $T_{dq}$ given in \eqref{tdq}. Since---with a slightly different scaling factor---this transformation was first introduced by Robert H. Park in 1929 \cite{park29} it is also often called Park transformation \citep[Appendix A]{teodorescu11}.
\endrem

\subsection{Instantaneous power}
\lab{ss:pow}

Power is one of the most important quantities in control, monitoring and operation of electrical networks. 
The first theoretical contributions to the definition of the power flows in an AC network date back to the early 20th century. However, these first definitions are restricted to sinusoidal steady-state conditions and based on the root-mean-square values of currents and voltages. As a consequence, these definitions of electric power are not well-suited for the purposes of network control under time-varying operating conditions \cite{akagi07}. 

The extension of the definition of electrical power to time-varying operating conditions is called ``instantaneous power theory'' in the power system and power electronics community \cite{akagi07,teodorescu11}. 
The development of this theory already begun in the 1930s with the study of active and non-active components of currents and voltages \cite{fryze32}. 
Among others, relevant contributions are \cite{buchholz50,depenbrock62,akagi83,willems92,depenbrock93,peng96,kim99}.

Today, it is widely agreed by researchers and practitioners \cite{willems92,peng96,teodorescu11} that the definitions of instantaneous power proposed in \cite{akagi83} and contained in \cite{akagi07} are well-suited for describing the power flows in three-phase three-wire systems and symmetric three-phase four-wire systems.
However, a proper definition of instantaneous power in asymmetric three-phase four-wire systems with nonzero neutral current and voltage is still an open (and controversial) field of research \cite{depenbrock04,akagi07,aredes09,teodorescu11}. A good overview of the research history on instantaneous power theory is given in \citep[Appendix B]{teodorescu11}.

Consider a symmetric three-phase voltage, respectively current, given by
\begequ
v_{abc}=\sqrt{2}V \begin{bmatrix} \sin(\theta)\\\sin(\theta-\frac{2\pi}{3})\\\sin(\theta+\frac{2\pi}{3}) \end{bmatrix}, \quad i_{abc}=\sqrt{2}I\begin{bmatrix} \sin(\varphi)\\\sin(\varphi-\frac{2\pi}{3})\\\sin(\varphi+\frac{2\pi}{3}) \end{bmatrix},
\lab{vi}
\endequ 
where $\theta :\R_{\geq0}\to \mathbb S$ and $\varphi :\R_{\geq0}\to \mathbb S$ are the phase angles and $\sqrt{2}V:\R_{\geq0}\to\R_{\geq 0},$ respectively $\sqrt{2}I:\R_{\geq0}\to\R_{\geq 0},$ the amplitudes of the respective three-phase signal. As shown in Section~\ref{ss:dq}, applying the transformation \eqref{tdq} to the signals given in \eqref{vi} yields
\begequ
\begs
v_{dq}&=\begin{bmatrix}V_d \\ V_q \end{bmatrix}=\sqrt{3}V\begin{bmatrix}\sin(\theta-\varrho)\\\cos(\theta-\varrho)
       \end{bmatrix},\\
i_{dq}&=\begin{bmatrix}I_d \\ I_q \end{bmatrix}=\sqrt{3}I\begin{bmatrix}\sin(\varphi-\varrho)\\\cos(\varphi-\varrho)
       \end{bmatrix}.
\end{split}
\lab{vidq}
\endequ
Based on the preceding discussion, the following definitions of instantaneous active, reactive and apparent power under symmetric, but not necessarily steady-state, conditions are used in this work. The definitions are based on \cite{akagi83,akagi07}, in which they are given in $\alpha\beta$-coordinates. For the purpose of the present paper, it is more convenient to equivalently define the instantaneous powers in $dq$-coordinates.

\begin{definition}
Let $v_{dq}(t)$ and $i_{dq}(t)$ be given by \eqref{vidq}. The instantaneous three-phase active power is defined as
\begequ
P(t):=v_{dq}^\top(t) i_{dq}(t)=V_d(t)I_d(t)+V_q(t)I_q(t).
\notag
\endequ
The instantaneous three-phase reactive power is defined as
\begequ
Q(t):=v_{dq}^\top(t) \begin{bmatrix}0 & 1\\-1&0 \end{bmatrix}i_{dq}(t)=V_d(t)I_q(t)-V_q(t)I_d(t).
\notag
\endequ
Finally, the instantaneous three-phase (complex) apparent power is defined as
\begequ
S(t):=P(t)+jQ(t).
\notag
\endequ
\lab{def:pow}
\end{definition}

From the above definition, straightforward calculations together with standard trigonometric identities yield
\begequ
\begs
P(t)&=3V(t)I(t)\cos(\theta(t)-\varphi(t)),\\
Q(t)&=3V(t)I(t)\sin(\theta(t)-\varphi(t)).
\end{split}
\notag
\endequ
It follows that whenever $v_{abc}$ and $i_{abc}$ given in \eqref{vi} possess constant amplitudes, as well as the same frequency, {\em i.e.}, $\dot\theta=\dot\varphi,$ all quantities $P,$ $Q$ and $S$ are constant. Moreover, then the given definitions of power are in accordance with the conventional definitions of power in a symmetric steady-state \cite{grainger94,akagi07,glover11}. 
For further information on definitions and physical interpretations of instantaneous power, also under asymmetric conditions, the reader is referred to \cite{willems92,peng96,emanuel04,willems05,akagi07,teodorescu11}. 

Since this work is mainly concerned with dynamics of generation units, all powers are expressed in ``Generator Convention'' \citep[Chapter 2]{glover11}. That is, delivered active power is positive, while absorbed active power is negative. Furthermore, capacitive reactive power is counted positively and inductive reactive power is counted negatively.

\subsection{Algebraic graph theory}
\lab{graphtheory}
An undirected graph of order $n$ is a tuple \mbox{$\mathcal G :=(\mathcal V, \mathcal E),$} where $\mathcal{V} := \{n_1, \ldots , n_n\}$ is the set of nodes and $\mathcal E \subseteq [\mathcal V]^2,$ 
$\mathcal E:=\{e_1,\ldots,e_m\},$ is the set of undirected edges, {\em i.e.}, the elements of $\mathcal E$ are subsets of $\mathcal V$ that contain two elements.
In the case of multi-agent systems, each node in the graph typically represents an individual agent.  For the purpose of the present work, an agent represents a DG or storage unit, respectively a load.  
The $l$-th edge connecting nodes $i$ and $k$ is denoted by $e_l=\{i,k\}.$
By associating an arbitrary ordering to the edges, the node-edge incidence
matrix $\mathcal B\in\R^{|\mathcal V|\times|\mathcal E|}$ is defined element wise as $b_{il} = 1,$ if node
$i$ is the source of the $l$-th edge $e_l,$ $b_{il} = -1,$ if $i$ is the sink of $e_l$ and $b_{il} = 0$ otherwise. For further information on graph theory, the reader is referred to, {\em e.g.}, \cite{diestel00,godsil01} and references therein.

\section{The microgrid concept}
\lab{sec:mg}
Microgrids have attracted a wide interest in different research and application communities over the last decade \cite{simpson12_2,hatziargyriou07,pogaku07,strbac15}.
However, the term ``microgrid'' is not uniformly defined in the literature \cite{lasseter02,hatziargyriou07,green07,katiraei08,chowdhury09,glover11,strbac15}.
Based on \cite{green07,hatziargyriou07,strbac15}, the following definition of an AC microgrid is employed in this survey paper. 

\begin{definition}
An AC electrical network is said to be an AC microgrid if it satisfies the following conditions.
\begenu
\item It is a connected subset of the LV or MV distribution system of an AC electrical power system.
\item It possesses a single point of connection to the remaining electrical power system. This point of connection is called point of common coupling (PCC).
\item It gathers a combination of generation units, loads and energy storage elements. 
\item It possesses enough generation and storage capacity to supply most of its loads autonomously during at least some period of time.
\item It can be operated either connected to the remaining electrical network or as an independent island network. The first operation mode is called grid-connected mode and the second operation mode is called islanded, stand-alone or autonomous mode.
\item In grid-connected mode, it behaves as a single controllable generator or load from the viewpoint of the remaining electrical system.
\item\lab{p5} In islanded mode, frequency, voltage and power can be actively controlled within the microgrid.
\endenu
\lab{def:mg}
\end{definition}

According to Definition~\ref{def:mg}, the main components in a microgrid are DG units, loads and energy storage elements. 
Typical DG units in microgrids are 
renewable DG units, such as photovoltaic (PV) units, wind turbines, fuel cells (FCs), as well as microturbines or reciprocating engines in combination with SGs. 
The latter two can either be powered with biofuels or fossil fuels \cite{lidula11,glover11}. 

Typical loads in a microgrid are residential, commercial and industrial loads \cite{lasseter02,katiraei08,lidula11}. It is also foreseen to categorize the loads in a microgrid with respect to their priorities, {\em e.g.}, critical and non-critical loads. This enables load shedding as a possible operation option in islanded mode \cite{lasseter02,lidula11}.

\begin{figure}[t]
	\centering 
\includegraphics[width=1\linewidth]{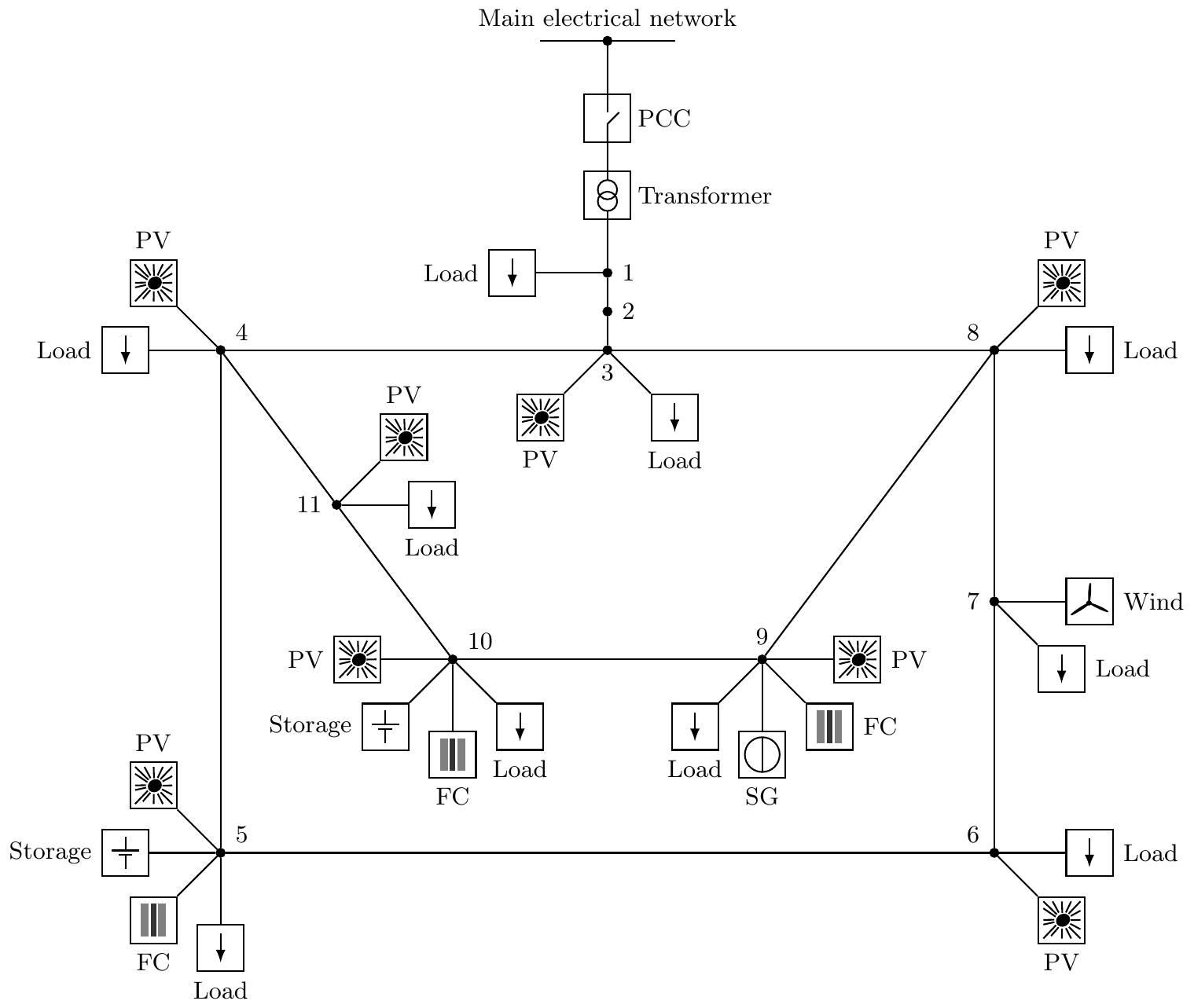}
\caption[Schematic representation of a microgrid]{Schematic representation of a microgrid. The microgrid is composed of several DG units, loads and storage devices. The DG units are inverter-interfaced photovoltaic (PV), fuel cell (FC) and wind power plants. In addition, a power generation unit is connected to the network via a synchronous generator (SG). The point of connection of the microgrid to the main network is called point of common coupling (PCC).}
\label{fig:mg}
\end{figure}

Finally, storage elements play a key-role in microgrid operation \cite{lidula11,glover11}. They are especially useful in balancing the power fluctuations of intermittent renewable sources and, hence, to contribute to network control. Possible storage elements are, {\em e.g.}, batteries, flywheels or supercapacitors. The combination of renewable DGs and storage elements is also an important assumption for the inverter models derived in this paper. An illustration of an exemplary microgrid is given in Fig.~\ref{fig:mg}.

Most of the named DG and storage units are either DC sources (PV, FC, batteries) or are often operated at variable or high-speed frequency (wind turbines, microturbines, flywheels). 
Therefore, they have to be connected to an AC network via AC or DC/AC inverters \cite{green07, teodorescu11}. For ease of notation, such devices are simply called ``inverters'' in the following.
Overviews on existing test-sites and experimental microgrids around the globe are provided in the survey papers \cite{hatziargyriou07,barnes07,lidula11,planas13,guo14}. 

\begrem
While not comprised in Definition~\ref{def:mg}, true island power systems are sometimes also called microgrids in the literature \cite{hatziargyriou07}. This can be justified by the fact that island power systems operating with a large share of renewable energy sources face similar technical challenges as microgrids. Nevertheless, an island power system differs from a microgrid in that it cannot be frequently connected to and disconnected from a larger electrical network \cite{green07}.
\endrem

\begrem
Microgrids can also be implemented as DC systems \cite{salomonsson08,salomonsson09,kwasinski11}. Definition~\ref{def:mg} can easily be adapted to this scenario by removing the property ``frequency control'' in point~\ref{p5}. Recent reviews of the main differences and challenges for AC and DC microgrids are given in \cite{wang12,planas13,justo13}.
\endrem

\section{Modeling of inverter-based microgrids}
\lab{sec4}

As outlined in the previous section, an AC microgrid is a spatially small power system the main components of which are renewable DG units, loads and energy storage elements interconnected through a network of AC transmission lines and transformers. 
Furthermore, most renewable DG and storage units are interfaced to the network via inverters.
As a consequence, fundamental network control actions, such as frequency or voltage control, have to be performed by inverter-interfaced units. 
This fact represents a fundamental difference to the operation of conventional power systems, where mainly SG units are responsible for network control.
Therefore and since the modeling of SGs is a well-covered topic in the literature \cite{kundur94,anderson02,machowski08}, we focus in the following on microgrids with purely inverter-interfaced DG and storage units. Also, it is straightforward to incorporate SG-based units into the microgrid model presented hereafter. 

In line with these considerations, an inverter-based microgrid can be represented by an undirected graph $\mathcal{G}=(\mathcal{N},\mathcal{E})$, in which---similarly to \cite{fiaz13}---nodes represent voltage buses, edges represent dynamic power lines and the topology of the network is fully described by the incidence matrix $\mathcal{B}$. The set of neighbors of node $i$ is denoted by $\mathcal N_i$ and contains all $k$ for which $e_l=\{i,k\}\in\mathcal E.$ Please see Section~\ref{graphtheory} for a brief introduction to algebraic graph theory. In the present case, we further assume that the set of nodes $\mathcal{N}$ can be partitioned into two subsets $\mathcal{N}_I$ and $\mathcal{N}_R$, associated to inverter and load nodes respectively. We next proceed as follows. First, we provide a description of the basic functionality and common operation modes of inverters that are instrumental for the modeling. Thereafter, we present models of inverters---depending on their mode of operation---loads, power lines and transformers. The section is concluded by combining the individual models to an overall 
representation of an inverter-based microgrid. To enhance readability, the subindex $i$, preceded by a comma when necessary, denotes in the sequel the elements corresponding to the $i$-th subsystem. 

\subsection{Basic functionality and common operation modes of inverters}
\begin{figure}
\centering
\includegraphics[width=1\linewidth]{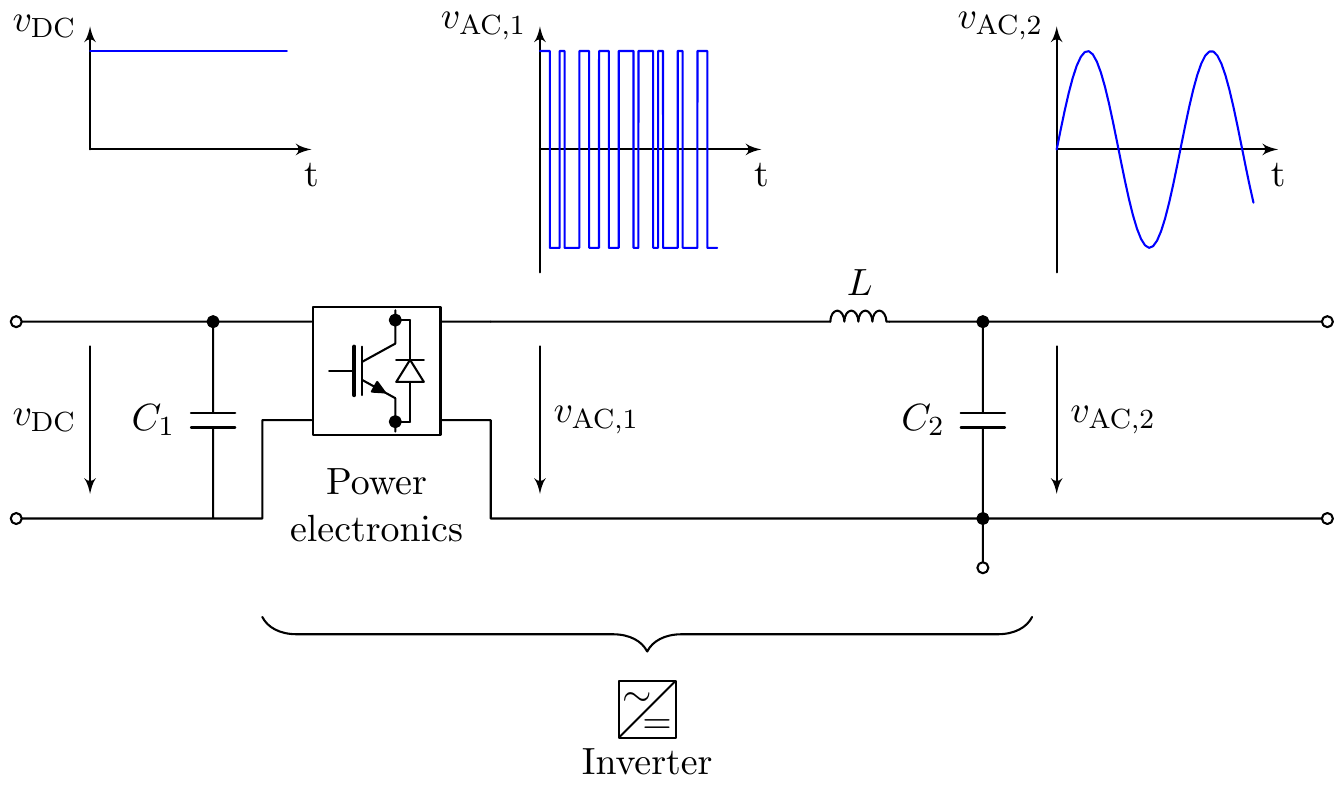}
\caption[Schematic representation of DC-AC conversion by an inverter]{Schematic representation of a DC-AC voltage conversion by a DC/AC inverter. The DC signal $v_{\text{DC}}:\R_{\geq0}\to\R$ on the left side is converted into an AC signal via power semiconductor devices. The generated AC signal $v_{\text{AC},1}:\R_{\geq0}\to\R^3$ at the output of the power electronics is not sinusoidal. Therefore, an $LC$ filter is connected in series with the power electronics to obtain an approximately sinusoidal ouput voltage $v_{\text{AC},2}:\R_{\geq0}\to\R^3$ with low harmonic content.}
\label{fig:DCAC}
\end{figure}

\begin{figure}
\centering
\includegraphics[width=1\linewidth]{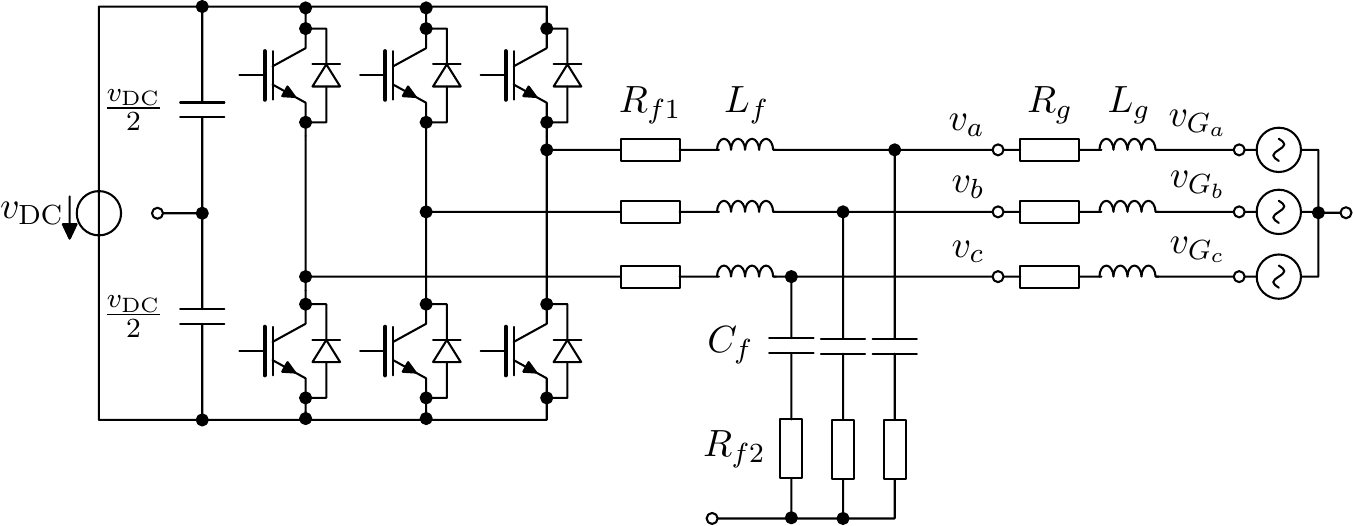}
\caption[Typical circuit of a two-level three-phase inverter with output filter]{Typical circuit of a two-level three-phase inverter with $LC$ output filter to convert a DC into a three-phase AC voltage. The inverter is constructed with insulated-gate bipolar transistors (IGBTs) and antiparallel diodes. The DC voltage is denoted by $v_{\text{DC}}:\R_{\geq0}\to\R,$ the three-phase AC voltage generated by the inverter with $v_{abc}:\R_{\geq0}\to\R^3,$ $v_{abc}=\col(v_{a},v_{b},v_{c})$ and the three-phase grid-side AC voltage by $v_G:\R_{\geq0}\to\R^3,$ $v_G=\col(v_{G_a}v_{G_b},v_{G_c}).$
The components of the output filter are an inductance $L_f\in\R_{>0},$ a capacitance $C_f\in\R_{>0}$ and two resistances $R_{f_1}\in\R_{>0},$ respectively $R_{f_2}\in\R_{>0}.$ 
Typically, the resistance $R_g\in\R_{>0}$ and the inductance $L_g\in\R_{>0}$ represent a transformer or an output impedance.
At the open connectors denoted by ``\textbf{o}`` the circuit can be grounded if desired.}
\label{fig:pwm}
\end{figure}

\lab{subsec:funcinverter}
Recall that 
inverters are key components of microgrids. Therefore, this section is dedicated to the model derivation of an inverter in a microgrid.
The basic functionality of an inverter is illustrated in Fig.~\ref{fig:DCAC}.
The main elements of inverters are power semiconductor devices \cite{erickson01,mohan07}. An exemplary basic hardware topology of the electric circuit of a two-level three-phase inverter constructed with insulated-gate bipolar transistors (IGBTs) and antiparallel diodes is shown in Fig.~\ref{fig:pwm}.
The conversion process from DC to AC is usually achieved by adjusting the on- and off-times of the transistors. These on- and off-time sequences are typically determined via a modulation technique, such as pulse-width-modulation \cite{erickson01,mohan07}. 
To improve the quality of the AC waveform, {\em e.g.}, to reduce the harmonics, the generated AC signal is typically processed through a low-pass filter constructed with $LC(L)$ elements.
Further information on the hardware design of inverters and related controls is given in \cite{erickson01,mohan07,zhong12}.

In microgrids, two main operation modes for inverters can be distinguished \cite{wang12,rocabert12}: grid-forming and grid-feeding mode. The latter is sometimes also called grid-following mode \cite{katiraei08} or PQ control \cite{lopes06}, whereas the first is also referred to as voltage source inverter (VSI) control \cite{lopes06}. 
The main characteristics of these two different operation modes are as follows \cite{lopes06,katiraei08,wang12,rocabert12}.

\begenu
\item Grid-forming mode (also: VSI control). 

\begin{figure*}
\centering
\subfloat[Schematic representation of an inverter operated in grid-forming mode based on \cite{rocabert12}. Bold lines represent electrical connections, while dashed lines represent signal connections. The current through the filter inductance is denoted by $i_{f,abc}:\R_{\geq0}\to\R^3$ and the inverter output voltage by $v_{abc}:\R_{\geq0}\to\R^3.$ Both quantities are fed back to a cascaded control consisting of an outer voltage and an inner current control. The reference signal $v_\text{ref}:\R_{\geq0}\to\R^3$ for the voltage controller is set by the designer, respectively a higher-level control. The IGBTs of the inverter are then controlled via signals generated by a modulator. The control structure can also be reduced to a pure voltage control.]
{\includegraphics[width=0.48\linewidth]{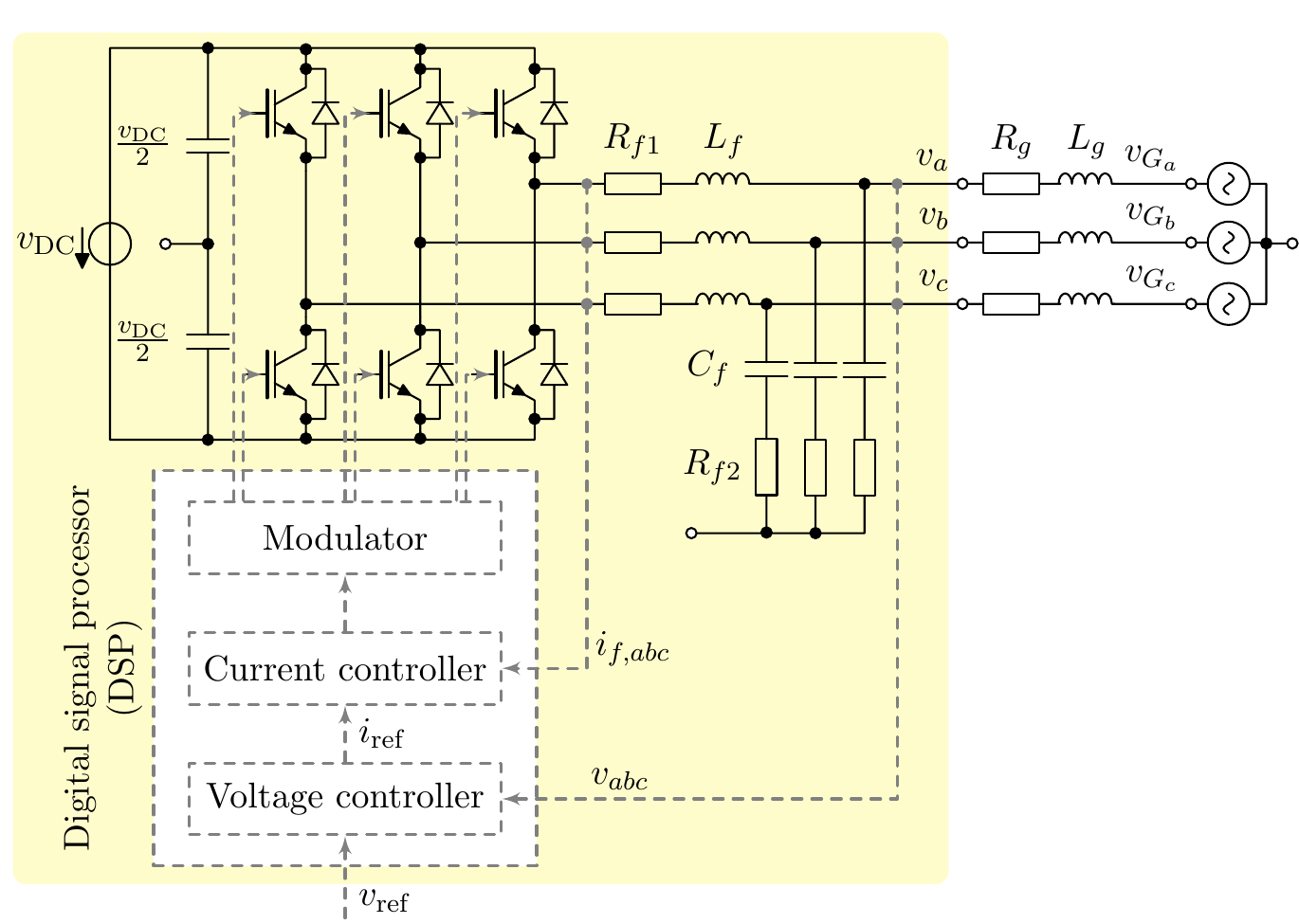}
\label{fig:gridFormCom}
}
\hfill
\subfloat[Schematic representation of an inverter operated in grid-feeding mode based on \cite{rocabert12}. Bold lines represent electrical connections, while dashed lines represent signal connections. As in Fig.~\ref{fig:gridFormCom}, the current through the filter inductance is denoted by $i_{f,abc}:\R_{\geq0}\to\R^3$ and the inverter output voltage by $v_{abc}:\R_{\geq0}\to\R^3.$ In grid-feeding mode, both quantities are fed back to a cascaded control consisting of an outer power and an inner current controller. The reference active and reactive powers $P_\text{ref}\in\R,$ respectively $Q_\text{ref}\in\R,$ are set by the designer or a higher-level control.]
{\includegraphics[width=0.48\linewidth]{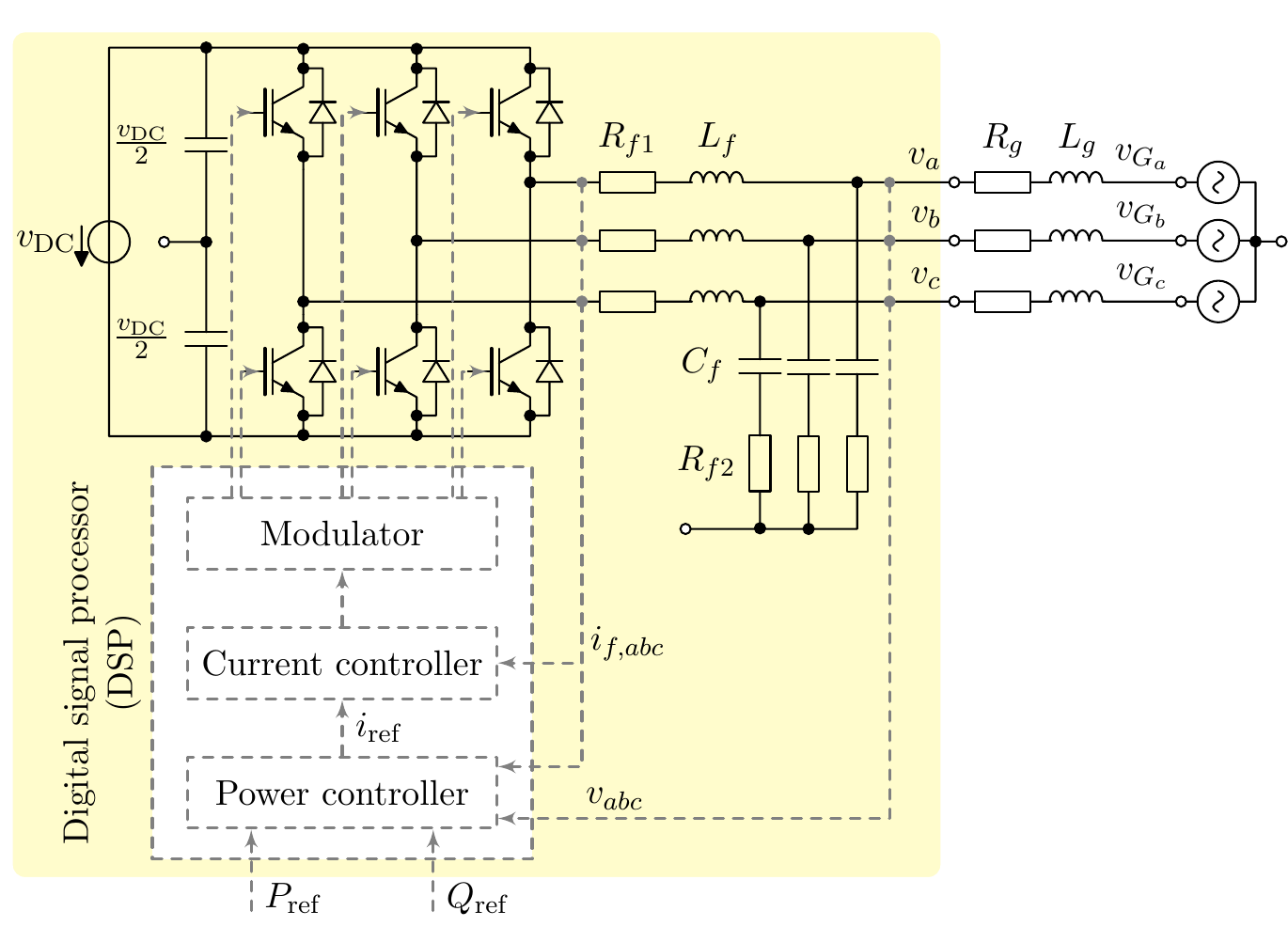}
\label{fig:gridFeed}
}
\caption{Schematic representation of the two main operation modes of inverters in microgrid applications: grid-forming and grid-feeding mode.}
\end{figure*}

The inverter is controlled in such way that its output voltage can be specified by the designer. This is typically achieved via a cascaded control scheme consisting of an inner current control and an outer voltage control as shown in Fig.~\ref{fig:gridFormCom}, based on \cite{rocabert12}. 
The feedback signal of the current control loop is the current through the filter inductance, while the feedback signal of the voltage control loop is the inverter output voltage $v_{abc}:\R_{\geq 0}\to\R^3$. The inner loop of the control cascade is not necessary to control the output voltage of the inverter and can hence also be omitted. Nevertheless, it is often included to improve the control performance.

\item Grid-feeding mode (also: grid-following mode, PQ control). 

The inverter is operated as power source, {\em i.e.}, it provides a pre-specified amount of active and reactive power to the grid. The active and reactive power setpoints are typically provided by a higher-level control or energy management system, see \cite{rocabert12,bolognani13, hans14}. 
Also in this case, a cascaded control scheme is usually implemented to achieve the desired closed-loop behavior of the inverter, as illustrated in Fig.~\ref{fig:gridFeed}. As in the case of a grid-forming inverter, the inner control loop is a current control the feedback signal of which is the current through the filter inductance. However, the outer control loop is not a voltage, but rather a power (or, sometimes, a current) control. The feedback signals of the power control are the active and reactive power provided by the inverter.

\endenu

In both aforementioned operation modes, the current and voltage control loops are, in general, designed with the objectives of rejecting high frequency disturbances, enhancing the damping of the output $LC(L)$ filter and providing harmonic compensation \cite{prodanovic03,blaabjerg06,mohamed08,pogaku07}.
Furthermore, nowadays, most inverter-based DG units, such as PV or wind plants, are operated in grid-feeding mode \cite{rocabert12}.
However, grid-forming units are essential components in AC power systems, since they 
are responsible for frequency and voltage regulation in the network. 
Therefore, in microgrids with a large share of renewable inverter-based DG units, grid-forming capabilities often also have to be provided by inverter-interfaced sources \cite{lopes06,katiraei08}.

\begrem
Some authors \cite{wang12,rocabert12} also introduce a third operation mode for inverters called grid-supporting mode.
Nevertheless, this last category is not necessary to classify typical operation modes of inverters in microgrids in the context of this work, since grid-supporting inverters are grid-forming inverters equipped with an additional outer control-loop to determine the reference output voltage. Such outer control-loops are, {\em e.g.}, the usual droop controls \cite{chandorkar93,guerrero13}. Therefore, the term ``grid-supporting inverter'' is not used in the following. 
\endrem

\begrem
In addition to the two control schemes introduced above, there also exist other approaches to operate inverters in microgrid applications. For example, \cite{beck07,visscher08,zhong11} propose to design the inverter control based on the model of an SG with the aim of making the inverter mimic as closely as possible the behavior of an SG.
However, to the best of the authors' knowledge, these approaches are not as commonly used as the control schemes shown in Fig.~\ref{fig:gridFormCom} and Fig.~\ref{fig:gridFeed}.
\endrem

\subsection{Modeling of grid-forming inverters}
\lab{subsec:gridform}
A suitable model of a grid-forming inverter for the purpose of control design and stability analysis of microgrids is derived. There are many control schemes available to operate an inverter in grid-forming mode, such as PI control in $dq$-coordinates \cite{pogaku07}, proportional resonant control \cite{fukuda01,teodorescu06} or repetitive control \cite{weiss04,hornik11} among others. An overview of the most common control schemes with an emphasis on $H_\infty$ repetitive control is given in \cite{zhong12}. For a comparison of different control schemes, the reader is referred to \cite{loh05}. The assumption below is key for the subsequent model derivation.

\begin{assumption}
Whenever an inverter 
operated in grid-forming mode
connects a fluctuating renewable generation source,
it is equipped with a fast-reacting storage.
\label{ass:stor}
\end{assumption}

Assumption~\ref{ass:stor} implies that the inverter can increase and decrease its power output within a certain range. This is necessary if the inverter should be capable of providing a fully controllable voltage also when interfacing a fluctuating renewable DG unit to the network. Furthermore, since the storage element is assumed to be fast-reacting, the DC-side dynamics can be neglected in the model. The capacity of the required DC storage element depends on the specific source at hand. Generally, the standard capacitive elements of an inverter don't provide sufficient energy storage capacity and an additional storage component, {\em e.g.}, a battery or flywheel, is required if an inverter is operated in grid-forming mode \cite{teodorescu11} See \cite{carrasco06} for a survey of energy storage technologies in the context of power electronic systems and renewable energy sources.

Due to the large variety of available control schemes, it is difficult to determine a standard closed-loop model of an inverter operated in grid-forming mode together with its inner control and output filter.
Therefore, the approach taken in this work is to represent such a system as a generic dynamical system. Note that the operation of the IGBTs of an inverter occurs typically at very high switching frequencies (2-20 kHz) compared to the network frequency (45-65 Hz). It is therefore common practice \cite{lopes06,green07,pogaku07,mohamed08,chini10} to model an inverter in network studies with continuous dynamics by using the so-called averaged switch modeling technique \cite{erickson01,chini10}, {\em i.e.}, by averaging the internal inverter voltage and current over a suitably chosen time interval such as one switching period.

It is convenient to partition the set $\mathcal{N}_I$ into two subsets, {\em i.e.}, $\mathcal{N}_I=\mathcal{N}_1\cup \mathcal{N}_2,$ such that $\mathcal{N}_1$ contains all nodes associated to grid-forming inverters and $\mathcal{N}_2$ contains those associated to grid-feeding inverters.
Consider a grid-forming inverter located at the $i$-th node of a given microgrid, {\em i.e.}, $i\in \mathcal{N}_1$. Denote its three-phase symmetric output voltage by $v_{abc,i}:\R_{\geq 0}\to\R^3$ with phase angle $\alpha_i:\R_{\geq 0}\to\mathbb S$ and amplitude $\sqrt{\frac{2}{3}}V_i:\R_{\geq 0}\to\R_{\geq 0}$, {\em i.e.},
\begequ
v_{abc,i}=\sqrt{\frac{2}{3}}V_i \begin{bmatrix} \sin(\alpha_i)\\\sin(\alpha_i-\frac{2}{3}\pi)\\\sin(\alpha_i+\frac{2}{3}\pi) \end{bmatrix}.\notag
\endequ
Furthermore, denote by $\omega_i:=\dot\alpha_i$ the frequency of the voltage $v_{abc,i}.$
Denote the state signal of the inverter with its inner control and output filter by $x_{i}:\R_{\geq 0}\to\R^m,$ its input signal by $v_{\text{ref},i}:\R_{\geq 0}\to\R^3$ and its interconnection port signals by $v_{abc,i}$ and $i_{abc,i}:\R_{\geq 0}\to\R^3,$ see Fig.~\ref{fig:gridFormCom}. 
Let $f_i:\R^m\times \R^3\times\R^3\to\R^m$ and $h_i:\R^m\times\R^3\to\R^3$ denote continuously differentiable functions and $\nu_i$ denote a nonnegative real constant. 
Then, the closed-loop inverter dynamics with inner control and output filter can be represented in a generic manner as
\begequ
\begin{split}
\nu_i\gamma_i\dot x_i&= f_i(x_i,v_{\text{ref},i},i_{abc,i}),\\
v_{abc,i}&=h_i(x_i,v_{\text{ref},i}),
\end{split}
\lab{invFull1}
\endequ
where the positive real constant $\gamma_i$ denotes the time-drift due to the clock drift of the processor used to operate the inverter, see \cite{schiffer15} for further details. Note that $i_{abc,i}$ represents a disturbance for the inner control system. 

One key objective in microgrid applications is to design suitable higher-level controls to provide a reference voltage $v_{\text{ref},i}$ for the system \eqref{invFull1} \cite{guerrero13}. Within the hierarchical control scheme discussed, {\em e.g.}, in \cite{guerrero11,guerrero13} this next higher control level corresponds to the primary control layer of a microgrid.
Let $z_i:\R_{\geq 0}\to\R^p$ denote the state signal of this higher-level control system, $u_{i}:\R_{\geq 0}\to\R^q$ its input signal and $v_{\text{ref},i}$ its output signal. Furthermore, let $g_i:\R^p\times\R^q\to\R^p$ and $w_i:\R^p\times\R^q\to\R^3$ be continuously differentiable functions.
Then, the outer control system of the inverter can be described by
\begequ
\begin{split}
\gamma_i\dot z_i&= g_i(z_i,u_i),\\
v_{\text{ref},i}&=w_i(z_i,u_i).
\end{split}
\lab{invFull2}
\endequ
Combining \eqref{invFull1} and \eqref{invFull2} yields the overall inverter dynamics for the $i$-th node, $i\in \mathcal{N}_1,$
\begequ
\boxed{
\begin{split}
\gamma_i\dot z_{i}&= g_i(z_{i},u_{i}),\\
\nu_i\gamma_i\dot x_{i}&= f_i(x_{i},w_i(z_{i},u_{i}),i_{abc,i}),\\
v_{abc,i}&=h_i(x_{i},w_i(z_{i},u_{i})).
\end{split}
}
\lab{invFull3}
\endequ

\subsection{Modeling of grid-feeding inverters and loads}
As discussed in Section~\ref{subsec:funcinverter}, grid-feeding inverters are typically operated as current or power sources. 
In order to achieve such behavior, the control methods employed to design the inner control loops of grid-forming inverters (see Section~\ref{subsec:gridform}) can equivalently be applied to operate inverters in grid-feeding mode. 
The current or power reference values are typically provided by a higher-level control, {\em e.g.}, a maximum power point tracker (MPPT) \cite{rocabert12}. 

We define the set $\mathcal{N}_\ell:= \mathcal{N}_2\cup\mathcal{N}_R$ that contains the nodes associated to grid-feeding inverters and loads.
As done for the model of a grid-forming inverter in \eqref{invFull3}, let $x_{\ell,i}:\R_{\geq 0}\to\R^r$ denote the state signal, $v_{abc,i}:\R_{\geq 0}\to\R^3$ and $i_{abc,i}:\R_{\geq 0}\to\R^3$ denote the interconnection port signals, $f_{\ell,i}:\R^r\times \R^3\to\R^r$ and $h_{\ell,i}:\R^r\times\R^3\to\R^3$ denote continuously differentiable functions and $\kappa_i$ denote a nonnegative real constant. We assume then a generic dynamic model of the form
\begequ
\boxed{
\begs
\kappa_i \dot x_{\ell,i}&=f_{\ell,i}(x_{\ell,i},i_{abc,i}),\\
v_{abc,i}&=h_{\ell,i}(x_{\ell,i},i_{abc,i}),
\end{split}
}
\lab{lm}
\endequ
for any node $i\in\mathcal N_\ell.$ 
In addition to grid-feeding inverters, the model \eqref{lm} can equivalently represent impedance ({\em e.g.}, $R$ parallel to $L$), current- or power-con\-trol\-led loads. Furthermore, a large variety of other load behaviors can be modeled by \eqref{lm}. We refer the reader to \cite{kundur94,vancutsem98} for further details on load modeling.

\subsection{Modeling of power lines and transformers}
The main purpose of the present paper is to provide a structured modeling procedure for microgrids. 
For ease of presentation, we make the following assumption.

\begin{assumption}
All power lines and transformers can be represented by symmetric three-phase $RL$ elements. 
\lab{ass:rl}
\end{assumption}

In light of Assumption~\ref{ass:rl} and to ease presentation, we solely use the term power lines to refer to the network interconnections in the following sections.
Also, note that it is straightforward to extent the modeling approach presented hereafter to more detailed power line or transformer models, as well as to DG units interfaced to the network via SGs. 

Recall that the topology of a microgrid can be conveniently described by an undirected graph $\mathcal{G}=(\mathcal{N},\mathcal{E})$, where $\mathcal E$ denotes the set of power lines interconnecting the different network nodes $i\sim\mathcal N.$
We associate an arbitrary ordering to the power lines $e_{l}\sim\mathcal E.$ Likewise, we assign to each power line $e_{l}\in\mathcal E$ a three-phase line current $i_{L,l}:\R_{\geq0}\to\R^3.$
With Assumption~\ref{ass:rl}, the power line $e_{l}\in\mathcal E$ connecting a pair of nodes $\{i,k\}\in[\mathcal N]^2$ is symmetric, {\em i.e.}, each phase of the power line $e_{l}$ is composed of a constant ohmic resistance $R_{l}\in\R_{>0}$ in series with a constant inductance $L_{l}\in\R_{>0}$ and $R_l$ as well as $L_l$ have the same value for each phase.
Furthermore, the voltage drop across the line is given by 
$$v_{L,l}:=v_{abc,i}-v_{abc,k}.$$
We denote the state of the $l$-th line by $x_{L,l}:=i_{L,l}$ and its interconnection port variables by $v_{L,l}$ and $i_{L,l}.$ Then, the model of the $l$-th line $e_{l}\in\mathcal{E}$ is given by
\begin{equation}
\begin{aligned}
L_{l}\dot{x}_{L,l}&=- R_{l} x_{L,l}+v_{L,l},\\
i_{L,l}&=x_{L,l}.
\lab{vk}
\end{aligned}
\end{equation}
For a compact derivation of the network dynamics, it is convenient to define the aggregated nodal voltages and currents
\begequ
v_{abc}:=\col(v_{abc,i})\in\R^{3|\mathcal N|},\;i_{abc}:=\col(i_{abc,i})\in\R^{3|\mathcal N|},\notag
\endequ
the aggregated line voltages and currents 
\begin{equation*}
v_{L}:=\col(v_{L,l})\in\R^{3|\mathcal E|},\quad i_{L}:=\col(i_{L,l})\in\R^{3|\mathcal E|},
\end{equation*}
as well as the matrices
\begequ
\begs
L:=\mathrm{diag}(L_{l})\in\R^{|\mathcal E|\times |\mathcal E|},\quad
R:=\mathrm{diag}(R_{l})\in\R^{|\mathcal E|\times |\mathcal E|}.
\end{split}
\notag
\endequ
Then, the three-phase interconnection laws can be obtained by following the approach used in \cite{fiaz13}, where Kirchhoff's current and
voltage laws (KCL and KVL) are expressed in relation to the node-edge incidence matrix $\mathcal{B},$ {\em i.e.},
\begin{equation}
i_{abc}=\mathcal{B}\otimes\mathbf{I}_3i_{L},\qquad \mathcal{B}^\top \otimes\mathbf{I}_3v_{abc}= v_{L}.
\lab{kcl}
\end{equation}
Hence, by combining \eqref{vk} with \eqref{kcl} the dynamical system representing the network is given by
\begequ
\begs
L\otimes\mathbf{I}_3\dot{x}_{L}&=- R\otimes\mathbf{I}_3x_{L}+\mathcal B^\top\otimes\mathbf{I}_3  v_{abc},\\
i_{abc}&=\mathcal{B}\otimes\mathbf{I}_3x_{L}.
\end{split}
\lab{iv1}
\endequ

We next transform the model \eqref{iv1} into $dq$-coordinates by means of the transformation $T_{dq}$ introduced in \eqref{tdq}. This coordinate transformation is instrumental for the model reduction carried out in Section~\ref{sec5}.
Let 
\begequ
\phi:= \mbox{mod}_{2\pi}\left(\omega^{\text{com}} t\right)\in\mathbb S,
\lab{phi}
\endequ
where the operator\footnote{The operator $\mbox {mod}_{2\pi}:\R \to [0, 2\pi)$, is defined as follows: $y=\mbox {mod}_{2\pi}\{x\}$ yields $y= x - k2\pi$ for some integer $k,$ such that $y\in [0, 2\pi)$.} $\mbox {mod}_{2\pi}$ is added to respect the topology of the torus.
Applying the transformation $T_{dq}$ with transformation angle $\phi$ to the signals $v_{abc,i}$ and $i_{abc,i},$ $i\sim\mathcal N,$ gives
\begequ
\hat v_{dq,i}:=T_{dq}(\phi)v_{abc,i}=\begin{bmatrix} \hat V_{d,i} \\ \hat V_{q,i}\end{bmatrix},\,\hat  i_{dq,i}:=T_{dq}(\phi)i_{abc,i}=\begin{bmatrix} \hat I_{d,i} \\ \hat I_{q,i}\end{bmatrix},\notag
\endequ
where the superscript "$\,\hat\cdot\,$" is introduced to denote signals in $dq$-coordinates with respect to the angle $\phi.$ This notation is used in the subsequent section, where a reduced-order model of a microgrid is derived by using several $dq$-transformation angles. Furthermore, following standard notation in power systems, the constant $\dot \phi=\omega^{\text{com}}$ is referred to as the rotational speed of the {\em common} reference frame.
Likewise, the signal $x_{L,l}$ in \eqref{vk} becomes
\begequ
\hat  x_{L,dq,l}:=T_{dq}(\phi)x_{L,abc,l}=\begin{bmatrix} \hat X_{L,d,l} \\ \hat X_{L,q,l}\end{bmatrix}.\notag
\endequ
Note that
\begequ
\begs
\hat{\dot{x}}_{L,dq,l}&=\dot{T}_{dq}(\phi) x_{L,l}+T_{dq}(\phi)\dot{x}_{L,l}\\
&=\omega^{\text{com}}\begin{bmatrix} 
                                         -\hat X_{L,q,l} \\ \hat X_{L,d,l} 
                                        \end{bmatrix}+T_{dq}(\phi)\dot{x}_{L,l}.\\
\end{split}
\notag
\endequ
Hence, \eqref{vk} reads in $dq$-coordinates as 
\begequ
\begin{split}
L_{l}\hat{\dot{x}}_{L,dq,l}&= L_l\left(\omega^{\text{com}}\begin{bmatrix} 
                                         -\hat X_{L,q,l} \\ \hat X_{L,d,l} 
                                        \end{bmatrix}+T_{dq}(\phi)\dot{x}_{L,l} \right)\\
&= - R_{l}\hat x_{L,dq,l}+L_{l}\omega^{\text{com}} \begin{bmatrix} 
                                         -\hat X_{L,q,l} \\ \hat X_{L,d,l}
                                        \end{bmatrix} +\hat v_{dq,l},\\
																				\hat i_{L,dq,l}&=\hat x_{L,dq,l}.
\end{split}
\lab{vkd}
\endequ
By defining the aggregated nodal voltages and currents in $dq$-coordinates
\begequ
\hat v_{dq}:=\col(\hat v_{dq,i})\in\R^{2|\mathcal N|},\;\hat i_{dq}:=\col(\hat i_{dq,i})\in\R^{2|\mathcal N|},\notag
\endequ
the aggregated line voltage and currents in $dq$-coordinates
\begin{equation*}
\hat v_{L,dq}:=\col(\hat v_{L,dq,l})\in\R^{2|\mathcal E|},\quad \hat x_{L,dq}:=\col(\hat i_{L,dq,l})\in\R^{2|\mathcal E|},
\end{equation*}
as well as the matrix
\begequ
\mathcal X:=\diag\left(L_{l}\omega^{\text{com}}\begin{bmatrix} 
                                        0 & -1 \\ 1 & 0
                                        \end{bmatrix}\right)\in\R^{2|\mathcal E|\times 2|\mathcal E|},\notag
\endequ
\eqref{iv1} becomes in $dq$-coordinates
\begequ
\boxed{
\begs
L\otimes\mathbf{I}_2\dot{\hat x}_{L,dq}&=(- R\otimes\mathbf{I}_2+\mathcal{X})\hat x_{L,dq}+\mathcal B^\top\otimes\mathbf{I}_2  \hat v_{dq},\\
\hat i_{dq}&=\mathcal{B}\otimes\mathbf{I}_2\hat x_{L,dq}.
\end{split}
}
\lab{ivdq}
\endequ

\subsection{Overall model }
\begin{figure}
\centering
\includegraphics[width=1\linewidth]{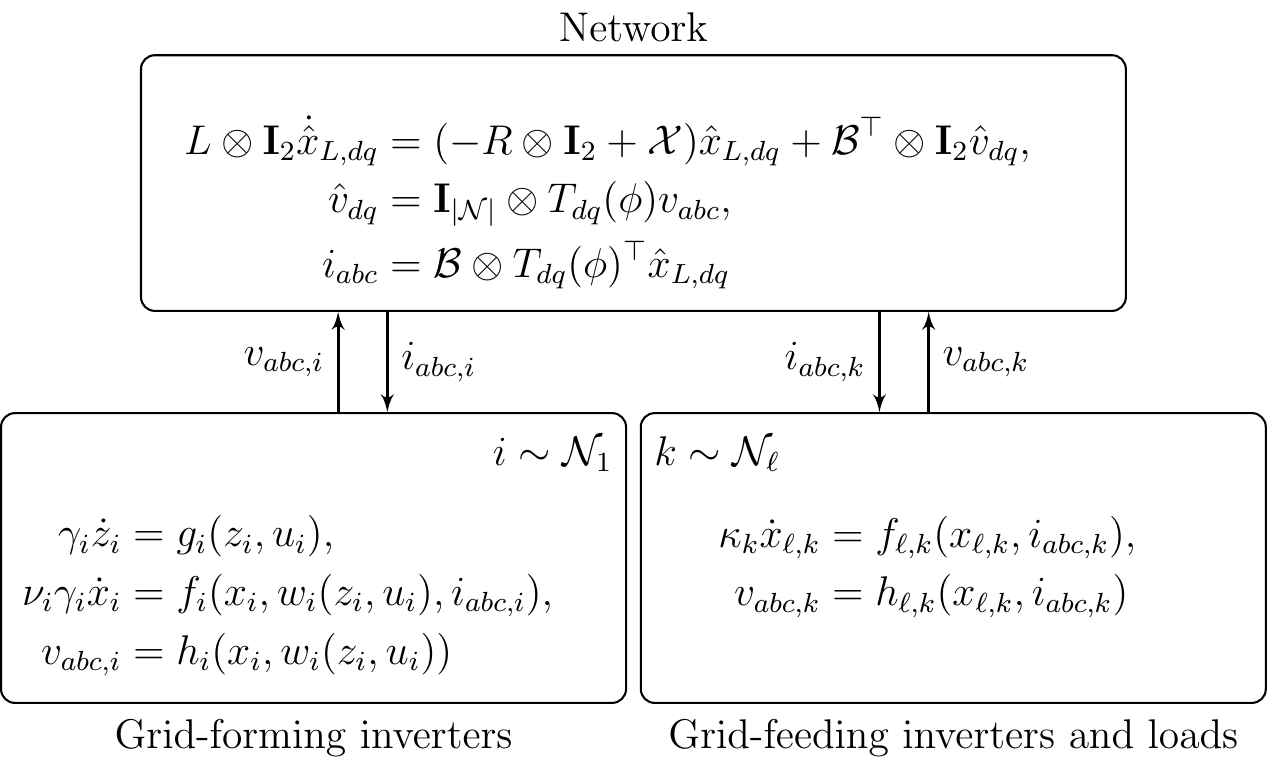}
\caption{Port-based representation of an inverter based-microgrid modeled by \eqref{overall}, \eqref{overall2}.
}
\label{fignet}
\end{figure}

By defining the state vectors $z\in\mathbb{R}^{p\vert \mathcal{N}_{1}\vert}$, $x\in\mathbb{R}^{m\vert \mathcal{N}_{1}\vert}$, $x_\ell\in\mathbb{R}^{r\vert \mathcal{N}_{\ell}\vert},$
the input $u\in\mathbb{R}^{q\vert \mathcal{N}_{1}\vert},$ the matrices
\begequ
\begs
\Gamma:&=\mathrm{diag}(\gamma_i)\in\R^{|\mathcal N_1|\times |\mathcal N_1|},\;\mathsf{V}:=\mathrm{diag}(\nu_i)\in\R^{|\mathcal N_1|\times |\mathcal N_1|},\\
 K:&=\mathrm{diag}(\kappa_i)\in\R^{|\mathcal N_2|\times |\mathcal N_2|},
\end{split}
\notag
\endequ
and combining \eqref{invFull3}, \eqref{lm} and \eqref{ivdq}, the overall microgrid model (see Fig. \ref{fignet}) is given by the differential equations
\begequ
\begs
\left[\Gamma\otimes\mathbf{I}_{p\vert \mathcal{N}_1\vert}\right]\dot z&=g(z,u),\\
\left[\mathsf{V}\Gamma\otimes\mathbf{I}_{m\vert \mathcal{N}_1\vert}\right]\dot x&=f(x,w(z,u),i_{abc}),\\
\left[K\otimes\mathbf{I}_{r\vert\mathcal{N}_{\ell}\vert}\right]\dot x_\ell&=f_\ell(x_\ell,i_{abc}),\\
\left[L\otimes\mathbf{I}_2\right]\dot{\hat x}_{L,dq}&=(- R\otimes\mathbf{I}_2+\mathcal{X})\hat x_{L,dq}+\mathcal B^\top\otimes\mathbf{I}_2  \hat v_{dq},
\end{split}
\label{overall}
\endequ
together with the algebraic relations
\begin{equation}
\begs
i_{abc}&=\mathcal{B}\otimes\mathbf{I}_3x_{L}= \mathcal{B}\otimes T_{dq}(\phi)^\top\hat x_{L,dq},\\
v_{abc,i}&=h_i(x_{i},w_i(z_{i},u_{i})),\\
 v_{abc,k}&=h_{\ell,k}(x_{\ell,k},i_{abc,k}),\\
\hat v_{dq}&=\mathbf{I}_{|\mathcal N|}\otimes T_{dq}(\phi)v_{abc},\quad i\sim \mathcal{N}_1,\quad k\sim\mathcal{N}_\ell.
\end{split}\lab{overall2}
\end{equation}

\section{To phasors and voltage sources via time-scale separation}
\label{sec5}

For the purpose of deriving an interconnected network model suitable for network control design and stability analysis, it is customary to make the following assumptions on \eqref{overall}, \eqref{overall2}, where $\epsilon$ stands for a generic small positive real constant.

\begin{assumption}
$\nu_i<\epsilon$ in \eqref{invFull3}, $i\sim\mathcal N_1.$ Therefore, $\dot x_i(t)=\underline 0_{m}$ for all $t\geq0.$ Furthermore, $v_{abc,i}=w_i(z_{i},u_{i}),$ $i\sim\mathcal N_1.$
\label{ass:vc}
\end{assumption}

\begin{assumption}
$\kappa_k<\epsilon$ in \eqref{lm}, $k\sim\mathcal{N}_\ell.$ Therefore, $\dot x_{\ell,k}(t)=\underline 0_{r}$ for all $t\geq0.$ 
Furthermore, the power balance at each node $k\in\mathcal{N}_\ell$ can be described by a ZIP model \cite{kundur94}, {\em i.e.},
\begequ
\begin{split}
P_k(\hat v_{dq,k},\hat i_{dq,k})&=-\left(a_{P,k}\|\hat v_{dq,k}\|_2^2+b_{P,k}\|\hat v_{dq,k}\|_2+c_{P,k}\right)\\
&:=P_k^*(\|\hat v_{dq,k}\|_2),\\
Q_k(\hat v_{dq,k},\hat i_{dq,k})&=-\left(a_{Q,k}\|\hat v_{dq,k}\|_2^2+b_{Q,k}\|\hat v_{dq,k}\|_2+c_{Q,k}\right)\\
&:=Q_k^*(\|\hat v_{dq,k}\|_2),
\end{split}\notag
\endequ
where $a_{P,k},$ $b_{P,k},$ $c_{P,k},$ $a_{Q,k},$ $b_{Q,k}$ and $c_{Q,k}$ are real constants and $P_k(\hat v_{dq,k},\hat i_{dq,k})$ and $Q_k(\hat v_{dq,k},\hat i_{dq,k})$ are calculated as given in Definition~\ref{def:pow}.
\lab{ass:load}
\end{assumption}

\begin{assumption}
$L<\epsilon\mathbf{I}_{|\mathcal E|}$ in \eqref{ivdq}. Therefore, \\$\dot {\hat x}_{L,dq}(t)=\underline 0_{2|\mathcal E|}$ for all $t\geq0.$
\lab{ass:di}
\end{assumption}

Assumption~\ref{ass:vc} is equivalent to the assumption that the inner current and voltage controllers track the voltage and current references instantaneously and exactly.
Usually, the current and voltage controllers in \eqref{invFull1} (see also Fig.~\ref{fig:gridFormCom}) are designed such that the resulting closed-loop system \eqref{invFull1} has a very large bandwidth compared to the control system located at the next higher control level represented by \eqref{invFull2} \cite{lopes06,coelho02,mohamed08}. 
If this time-scale separation is followed in the design of the system \eqref{invFull3}, the first part of Assumption~\ref{ass:vc} can be mathematically formalized by invoking singular perturbation theory \citep[Chapter 11]{khalil00}, \cite{kokotovic99}. The second part of Assumption~\ref{ass:vc} expresses the fact that the inner control system \eqref{invFull1} is assumed to track the reference $v_{\text{ref},i}=w_i(z_{i},u_{i})$ exactly, independently of the disturbance $i_{abc,i}.$ 
Typical values for the bandwidth of \eqref{invFull1} reported in \cite{mohamed08,pogaku07} are in the range of $400-600$ Hz, while those of \eqref{invFull2} are in the range of $2-10$ Hz. 

Assumption~\ref{ass:load} implies that the dynamics of loads and grid-feeding units can be neglected. This assumption is also frequently employed in microgrid and power system stability studies, where loads are often modeled as either constant impedance (Z), constant current (I) or constant power loads (P) or a combination of them (ZIP) \cite{kundur94,vancutsem98}. Similarly, grid-feeding units with positive active power injection are represented by setting $a_{P,k}=a_{Q_k}=0$ and $b_{P,k}$ or $c_{P,k}$ to negative values. The values for $b_{Q,k}$ and $c_{Q,k}$ should be chosen in dependency of the reactive power contribution of the unit.

Assumption~\ref{ass:di} is standard in power system analysis \cite{kundur94,grainger94,sauer98,anderson02,machowski08,glover11}. 
The usual justification of Assumption~\ref{ass:di} is that the line dynamics evolve on a much faster time-scale than the dynamics of the generation sources.
In the present case, Assumption~\ref{ass:di} is justified whenever Assumption~\ref{ass:vc} is employed, since the line dynamics \eqref{ivdq} are typically at least as fast as those of the internal inverter controls \eqref{invFull1}, see, {\em e.g.}, \cite{pogaku07}.
Again, Assumption~\ref{ass:di} can be mathematically formalized by invoking singular perturbation arguments \citep[Chapter 11]{khalil00}, \cite{kokotovic99}. 

Under Assumption~\ref{ass:vc}, the model of each grid-forming inverter \eqref{invFull3} reduces to
\begin{equation}
\begin{aligned}
\gamma_i\dot z_{i}&= g_i(z_{i},u_{i}),\\
v_{abc,i}&=w_i(z_{i},u_{i}), \quad i\sim \mathcal{N}_1.
\end{aligned}
\label{netr1}
\end{equation}
The model \eqref{netr1} represents the inverter as an AC voltage source, the amplitude and frequency of which can be defined by the designer. The system \eqref{netr1} is a very commonly used model of a grid-forming inverter in microgrid control design and analysis \cite{lopes06,green07,katiraei08,schiffer13_2}. 

Furthermore, often a particular structure of \eqref{netr1} is used in the literature \cite{simpson12_2,simpson13,schiffer_cdc12,schiffer13_2,ainsworth13,muenz14}.
As discussed in Section~\ref{ss:ac}, a symmetric three-phase voltage can be completely described by its phase angle and its amplitude. In addition, it is usually preferred to control the frequency of the inverter output voltage, instead of the phase angle.
Hence, a suitable model of the inverter at the $i$-th node is given by \cite{schiffer13_2,schiffer_cdc12}
\begequ
\label{invmod}
\boxed{
\begs
\gamma_i\dot{\alpha}_i&=\omega_i=u_{i}^{\delta}, \\
V_i&=u_i^{\text{V}},\\
v_{abc,i}&=v_{abc,i}(\alpha_i,V_i),
\end{split}
}
\end{equation}
where \mbox{$u_{i}^\delta:\R_{\geq0}\rightarrow \R$} and \mbox{$u_i^V:\R_{\geq0}\rightarrow \R$} are control signals.

Usually, it is also assumed that the active and reactive power output of the inverter is measured and processed through a filter to obtain the power components corresponding to the fundamental frequency \cite{pogaku07,coelho02,mohamed08}
\begequ
\label{tilpq}
\boxed{
\begs
\gamma_i\tau_{P_i}\dot P^m_i&=-{P}^m_i+P_i,\\
\gamma_i\tau_{P_i}\dot Q^m_i&=-{Q}^m_i+Q_i.
\end{split}
}
\endequ
Here, $P_i$ and $Q_i$ are the active and reactive power injections of the inverter,
$P_i^m:\R_{\geq0}\to \R$ and $Q_i^m:\R_{\geq0}\to \R$ their measured values and $\tau_{P_i}\in \R_{>0}$ is the time constant of the low pass filter. 

Note that whenever the particular form \eqref{invmod}, \eqref{tilpq} of \eqref{invFull2} is considered and the measured and filtered power signals are used as feedback signals in the controls $u_i^\delta,$ respectively $u_i^\text{V},$ then the bandwidth of the overall control system is limited by the bandwidth of the measurement filter. Hence, if $\tau_{P_i}\gg\nu_i,$ then Assumption~\ref{ass:vc} is justified. 

With Assumption~\ref{ass:load}, \eqref{lm} can be represented by the algebraic relation
\begin{equation}
\boxed{
\begin{aligned}
P_k(\hat v_{dq,k},\hat i_{dq,k})&=P_k^*(\|\hat v_{dq,k}\|_2),\\
Q_k(\hat v_{dq,k},\hat i_{dq,k})&=Q_k^*(\|\hat v_{dq,k}\|_2),\quad k\sim\mathcal N_\ell.
\end{aligned}
}
\label{netr2}
\end{equation}
Finally, under Assumption~\ref{ass:di}, the network model \eqref{ivdq} is also static and given by
\begequ
\boxed{
\hat i_{dq}=\mathcal B \otimes \mathbf{I}_2\left(R\otimes \mathbf{I}_2-\mathcal X\right)^{-1}\mathcal B^\top\otimes \mathbf{I}_2 \hat v_{dq}.
}
\lab{vkd2}
\endequ

\begin{figure}
\centering
\includegraphics[width=1\linewidth]{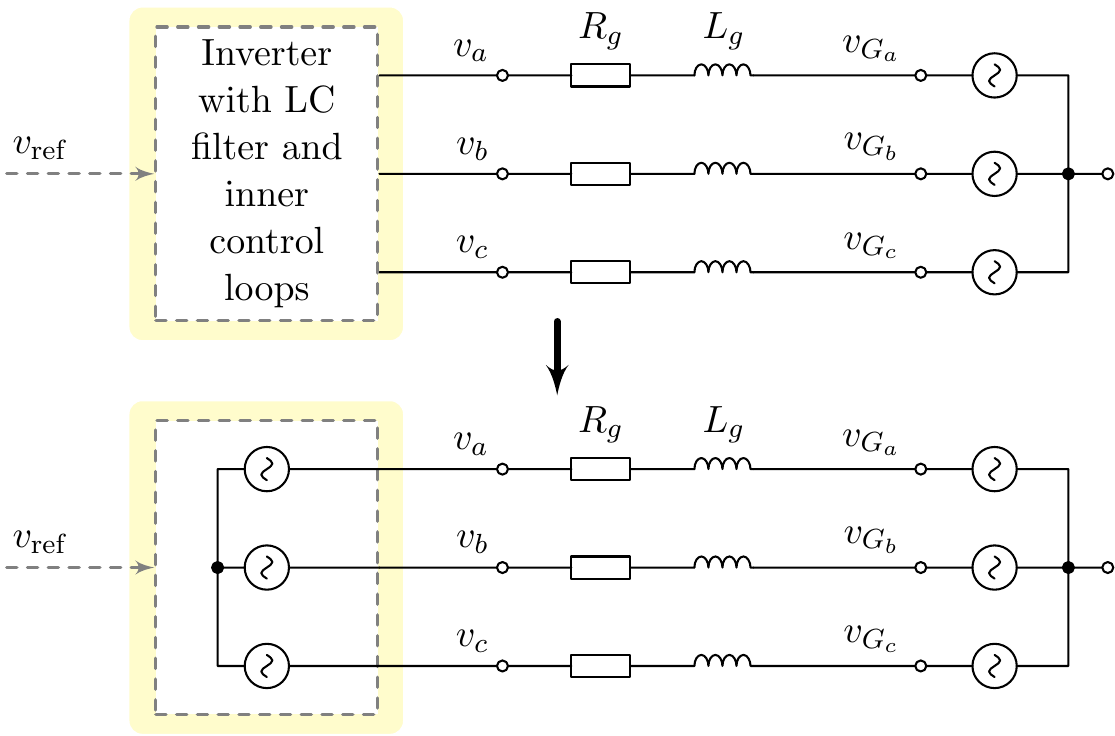}
\caption[Representation of an inverter operated in grid-forming mode as ideal controllable voltage source]{Simplified representation of an inverter operated in grid-forming mode as ideal controllable voltage source. Bold lines represent electrical connections, while dashed lines represent signal connections.
}
\label{fig:gridFormSim}
\end{figure}

\begin{figure}
\centering
\includegraphics[width=1\linewidth]{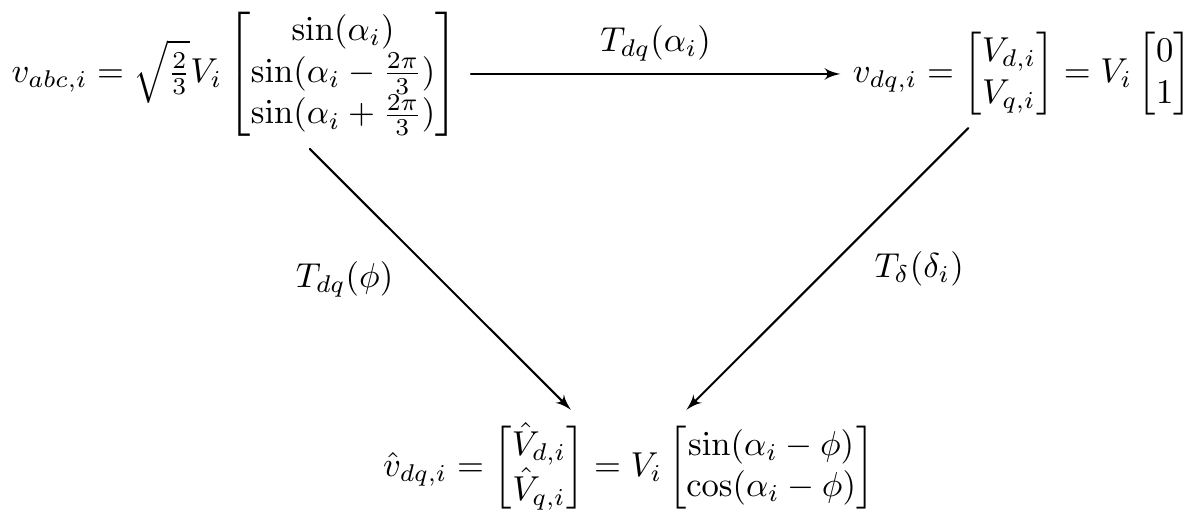}
\caption[Illustration of the different coordinate frames used to derive a model of an electrical network]
{Illustration of the different coordinate frames used to derive the model of an electrical network given in \eqref{ihYv}. The signal $v_{abc,i}:\R_{\geq0}\to\R^3$ denotes the three-phase voltage at the $i$-th bus with phase angle $\alpha_i:\R_{\geq0}\to\mathbb S$ and amplitude $V_i:\R_{\geq0}\to\R_{\geq 0},$ $i\sim\mathcal N$. 
The mappings $T_{dq}$ and $T_{\delta}$ are given in \eqref{tdq}, respectively \eqref{Ti}. The angle $\delta_i:\R_{\geq0}\to\mathbb S$ is defined in \eqref{dethe}. Note that, by construction, $\alpha_i-\delta_i=\mbox{mod}_{2\pi}\left(\omega^{\text{com}}t\right)=\phi,$ where the real constant $\omega^{\text{com}}$ denotes the speed of the common $dq$-reference frame.}
\label{fig:cF}
\end{figure}

The reduced-order model \eqref{invmod} - \eqref{vkd2} is still rather complex to handle, as the variables of \eqref{invmod} are expressed in $abc$-coordinates, while those of \eqref{netr2} and \eqref{vkd2} are expressed in common $dq$-coordinates.
Therefore, a more compact representation of \eqref{invmod} - \eqref{vkd2} is derived in the following. To this end, it is convenient to recall that $\alpha_i$ is the angle of the voltage at the $i$-th node with initial condition $\alpha_{0_i},$ $i\sim\mathcal N$ and to define
\begequ
\delta_i:=\alpha_{0_i}+\int_0^t (\dot\alpha_i -\omega^{\text{com}})d\tau\in\mathbb S,\quad i\sim \mathcal{N}.
\label{dethe}
\endequ
Let $\varpi:\R_{\geq 0}\to\mathbb S$ and consider the mapping $T_{\delta}:\mathbb S \rightarrow \R^{2\times 2},$ 
\begequ
T_{\delta}(\varpi):=\begin{bmatrix}
          \cos(\varpi) & \sin(\varpi) \\ -\sin(\varpi) & \cos(\varpi)
         \end{bmatrix}.
\label{Ti}         
\endequ
Note that, with $\delta_i$ defined in \eqref{dethe},
\begequ
\alpha_i-\delta_i= \mbox{mod}_{2\pi}\left(\omega^{\text{com}}t\right)= \phi,\quad i\sim \mathcal{N},\notag
\endequ
and that straightforward algebraic manipulations yield
\begequ
T_{dq}(\phi)= T_{\delta}(\delta_i)T_{dq}(\alpha_i).\notag
\endequ
Hence, by construction,
\begequ
\hat v_{dq,i}=T_{dq}(\phi)v_{abc,i}=T_{\delta}(\delta_i)T_{dq}(\alpha_i)v_{abc,i}=T_{\delta}(\delta_i)V_i\begin{bmatrix}0\\ 1 \end{bmatrix},
\lab{vdqdelta}
\endequ
which makes it convenient to define
\begequ
v_{dq,i}:=\begin{bmatrix}V_{d,i}\\ V_{q,i} \end{bmatrix}=V_i\begin{bmatrix}0\\ 1 \end{bmatrix}, \quad i\sim \mathcal{N}.
\lab{vdqloc}
\endequ
The variables $v_{dq,i}$ are referred to as {\em local} $dq$-coordinates of $v_{abc,i}$ in the following. 
The relation between $v_{abc,i},$ $\hat v_{dq,i}$ and $v_{dq,i}$ is illustrated in Fig.~\ref{fig:cF}.  

It is convenient to represent \eqref{vdqdelta} in the complex plane
\begequ
\hat V_{qd,i}:=\hat V_{q,i} +j \hat V_{d,i}= (\cos(\delta_i)+j\sin(\delta_i))V_{qd,i}= e^{j \delta_i}V_{qd,i}, 
\label{vdqh}
\endequ
where $V_{qd,i}=V_{q,i}+jV_{d,i},$ $i\sim\mathcal N.$ Equivalently, let
\begequ
\hat I_{qd,i}:=\hat I_{q,i} +j \hat I_{d,i}=  e^{j \delta_i}I_{qd,i} 
\label{idqh}
\endequ
and define
\begequ
\begs
\hat {V}_{qd}  :=&\hat V_q + j \hat V_d \in \mathbb C^{|\mathcal N|},\quad \hat {I}_{qd}  := \hat I_q + j \hat I_d \in \mathbb C^{|\mathcal N|},\\
{V}_{qd} := &V_q + j V_d \in \mathbb C^{|\mathcal N|},\quad {I}_{qd}  := I_q + j  I_d \in \mathbb C^{|\mathcal N|}.
\end{split}
\lab{ph2}
\endequ

Then, with $X:=\diag(X_l)=\diag(L_{l}\omega^{\text{com}})\in\R^{|\mathcal E|\times |\mathcal E|},$ we can rewrite \eqref{vkd2} as
\begequ
\hat I_{qd}=\mathcal B \left(R+j X\right)^{-1}\mathcal B^\top\hat V_{qd}.
\lab{vkd3}
\endequ

Note that the reactances $X_l=L_{l}\omega^{\text{com}}$ are calculated at the frequency $\omega^{\text{com}},$ which, under the made assumptions, should be chosen as the (constant) synchronous frequency of the network---denoted by $\omega^s\in\R$ in the following\footnote{Under the made assumptions, \eqref{vkd2} is the equilibrium of the "fast" line dynamics \eqref{vkd} \citep[Chapter 11]{khalil00}. Hence, in order for the currents $\hat i_{dq}$ and voltages $\hat v_{dq}$ to be constant in steady-state, $\omega^{\text{com}}$ has to be chosen identically to the synchronous steady-state network frequency.}. Typically, $\omega^s \in2\pi[45,65]$ rad/s. 

\begrem
The form \eqref{ph2} is a very popular representation and 
these complex quantities are often denoted as phasors \cite{anderson02, zhong12}. 
Furthermore, by using Euler's formula \cite{hazewinkel93}, \eqref{ph2} can also be rewritten in polar form.
Note, however, that, unlike, {\em e.g.}, \cite{anderson02, zhong12}, other authors define a phasor as a complex sinusoidal quantity with a constant frequency \cite{glover11}. 
\endrem

Define the admittance matrix of the electrical network by
\begequ
{\mathcal Y }:=\mathcal B \left(R+j X\right)^{-1}\mathcal B^\top\in \mathbb C^{|\mathcal N|\times |\mathcal N|}
\lab{mY}
\endequ
and
\begequ
\begs
G_{ii}:=&\Re(\mathcal Y_{ii}),\quad B_{ii}:=\Im(\mathcal Y_{ii}),\\
 Y_{ik}:=&G_{ik}+jB_{ik}:=-\mathcal Y_{ik},\quad i\neq k.
\end{split}
\notag
\endequ
Moreover, it follows immediately that
\begequ
\mathcal Y_{ik}=\begin{cases} 0 & \text{\begin{tabular}{l l}if nodes $i$ and $k$ are\\ not connected\end{tabular}} \\
-(R_l+jX_l)^{-1} &  \text{\begin{tabular}{l l}if nodes $i$ and $k$ are\\ connected by line $l$\end{tabular}} \end{cases}\notag
\endequ
and
\begequ
G_{ii}+jB_{ii}=\sum_{l\sim \mathcal E_i} (R_l+jX_l)^{-1},\notag
\endequ
where $\mathcal E_i$ denotes the set of edges associated to node $i.$ 
Inserting \eqref{vdqh} and \eqref{idqh} into \eqref{vkd3} yields
\begequ
I_{qd} = \diag\left(e^{-j\delta_i}\right)\mathcal{Y} \diag\left(e^{j\delta_i}\right) V_{qd}.
\lab{ihYv}
\endequ
Recall that $V_{qd}$ and $I_{qd}$ defined in \eqref{ph2} are expressed in local $dq$-coordinates.
By making use of \eqref{vdqloc} and \eqref{mY}, \eqref{ihYv} can be written component-wise as
\begequ
\begs
I_{qd,i}&=I_{q,i}+jI_{d,i},\\
I_{q,i}&=
 G_{ii} V_{i}-\sum_{k\sim\mathcal N_i} \left(G_{ik}\cos(\delta_{ik})+B_{ik}\sin(\delta_{ik})\right) V_{k},\\
 I_{d,i}&=
B_{ii} V_{i}-\sum_{k\sim\mathcal N_i}\left(B_{ik}\cos(\delta_{ik})-G_{ik}\sin(\delta_{ik})\right) V_{k},
\end{split}
\label{Idi}
\endequ
$i\sim\mathcal N,$ where, for ease of notation, angle differences are written as \mbox{$\delta_{ik}:=\delta_i-\delta_k.$}
Furthermore, from Definition~\ref{def:pow} together with \eqref{vdqloc} and \eqref{Idi}, the power flows in the network are given by
\begequ
\begs
&P_{i}= V_i I_{q,i}=\\
&G_{ii}V_{i}^2- \sum_{k\sim\mathcal N_i} \left( G_{ik} \cos (\delta_{ik}) + B_{ik}\sin (\delta_{ik}) \right) V_{k}V_{i},\\
&Q_{i} = -V_i I_{d,i}\\
&=-B_{ii}V_{i}^2+ \sum_{k\sim\mathcal N_i}\left( B_{ik} \cos (\delta_{ik}) - G_{ik}\sin (\delta_{ik}) \right) V_{k}V_{i}.
\end{split}
\label{powFlow}
\endequ
The equations \eqref{powFlow} are the standard power flow equations used in most recent work on microgrid control design and stability analysis, {\em e.g.}, \cite{simpson12_2,schiffer13_2,ainsworth13,muenz14}. 

\begrem
Note that for any other choices of the transformation angle in local $dq$-coordinates $V_{d_i}\neq0.$ This is usually the case when modeling SGs, since the angle of the internal machine electromagnetic force (EMF) is in general not known. Then, the equations \eqref{powFlow} become slightly more involved, see \citep[Chapter 9]{anderson02}.
\endrem

\begin{figure}
\centering
\includegraphics[width=1\linewidth]{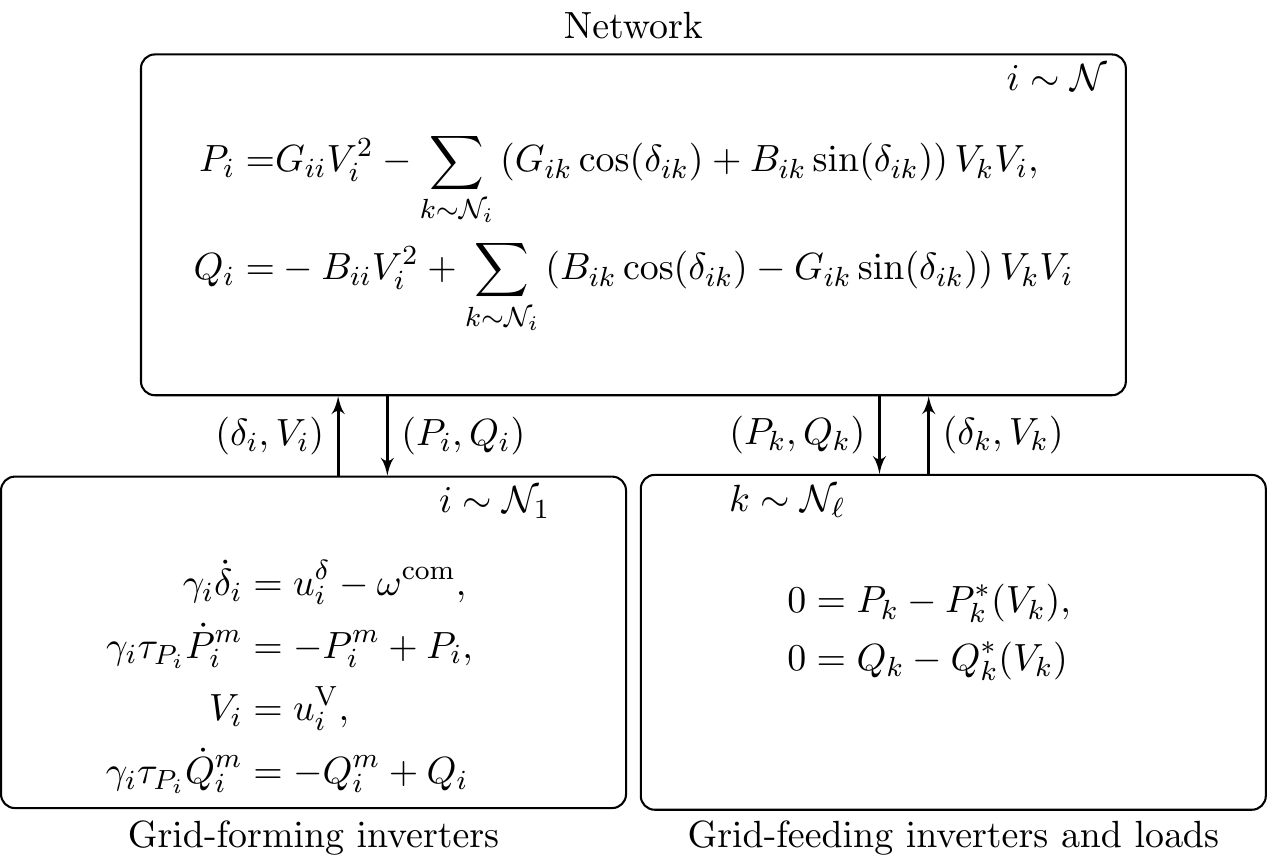}
\caption{Reduced microgrid model in standard representation with grid-forming inverters modeled by \eqref{inv}, as well as loads and grid-feeding inverters given by \eqref{netr2} with $\|\hat v_{dq,k}\|_2=V_k.$ The network is represented by the power flow equations \eqref{powFlow}. This compact model is obtained from the model \eqref{invmod} - \eqref{vkd2} by expressing the current and power flows in local $dq$-coordinates $v_{dq,i}$ and $i_{dq,i},$ $i\sim\mathcal N,$ see \eqref{ihYv} and Fig.~\ref{fig:cF}.}
\label{fignet3}
\end{figure}

Furthermore, in local $dq$-coordinates, the particular inverter model \eqref{invmod}, \eqref{tilpq}, is given by
\begequ
\begs
\gamma_i\dot{\delta_i}=\omega_i- \omega^{\text{com}}&= u_i^\delta- \omega^{\text{com}}, \\
\gamma_i\tau_{P_i}\dot P^m_i&=-{P}^m_i+P_i,\\
V_i&=u_i^V,\\
\gamma_i\tau_{P_i}\dot Q^m_i&=-{Q}^m_i+Q_i,
\end{split}
\label{inv}
\endequ
with $V_{qd,i}=V_i$ (see \eqref{vdqloc}) and $P_i$ and $Q_i$ given by \eqref{powFlow}. 
Finally, recall \eqref{netr2} and note that
$$
\|\hat v_{dq,k}\|_2=\|\hat V_{qd,k}\|_2=\|V_{qd,k}\|_2=V_k,\quad k\in\mathcal N_\ell.
$$
This completes the reformulation of the model \eqref{invmod} - \eqref{vkd2}.
The final overall microgrid model is given by \eqref{netr2}, \eqref{powFlow}, \eqref{inv} and shown in Fig.~\ref{fignet3}. This is the standard model employed throughout the literature.

The section is concluded by deriving a vector-based formulation of the microgrid model \eqref{netr2}, \eqref{powFlow}, \eqref{inv}. To this end, we define the vectors
\begequ
\begs
\delta_I:=&\col(\delta_i)\in\mathbb S^{| \mathcal{N}_1|},\;V_I:=\col(V_i)\in\R_{>0}^{| \mathcal{N}_1|},\\
u^\delta:=&\col(u_i^\delta)\in\R^{| \mathcal{N}_1|},\;u^V:=\col(u_i^V)\in\R^{| \mathcal{N}_1|},\\
P_I:=&\col(P_i) \in \R^{| \mathcal{N}_1|},\; Q_I:=\col(Q_i) \in \R^{| \mathcal{N}_1|},\\
P_\ell:=&\col(P_k) \in \R^{|\mathcal N_\ell|},\; Q_\ell:=\col(Q_k) \in \R^{|\mathcal N_\ell|},\\
P_\ell^*:=&\col(P_k^*(V_k)) \in \R^{|\mathcal N_\ell|},\; Q_\ell^*:=\col(Q_k^*(V_k)) \in \R^{|\mathcal N_\ell|},
\end{split}\notag
\endequ
with $P_i,$ $P_k,$ $Q_i,$ and $Q_k$ given by \eqref{powFlow}, as well as the matrix
$$
T:=\diag(\tau_{P_i}) \in \R^{|\mathcal N_1| \times | \mathcal{N}_1|}.
$$
Then the system \eqref{invmod} - \eqref{vkd2}
can be written equivalently by means of \eqref{netr2}, \eqref{powFlow}, \eqref{inv} as
\begequ
\begs
\Gamma\dot{\delta_I}&=u^\delta-\omega^{\text{com}}\mathds{1}_{|\mathcal N_1|}, \\
\Gamma T\dot P^m&=-{P}^m+P_I,\\
V_I&=u^V,\\
\Gamma T\dot Q^m&=-{Q}^m+Q_I,\\
\underline 0_{|\mathcal N_\ell|}&=P_\ell-P_\ell^*, \\
\underline 0_{|\mathcal N_\ell|}&=Q_\ell-Q_\ell^*,
\end{split}
\label{mgo}
\endequ
where the last $2|\mathcal N_\ell|$ algebraic equations correspond to the power balances at nodes $k\sim\mathcal N_\ell.$

This section has illustrated the main modeling steps and assumptions, which lead from the detailed microgrid model \eqref{overall}, \eqref{overall2} to the model \eqref{mgo}, \eqref{powFlow}, respectively \eqref{netr2}, \eqref{powFlow}, \eqref{inv}.
The model \eqref{mgo}, \eqref{powFlow} is frequently used in the analysis and control design of microgrids \cite{simpson12_2,simpson13,bouattour13_2,schiffer13,schiffer13_2,ainsworth13,muenz14,schiffer14,schiffer14_2}. Some of the mentioned work is conducted under additional assumptions such as instantaneous power measurements \cite{simpson12_2,ainsworth13,muenz14}, constant voltage amplitudes \cite{simpson12_2,bouattour13_2,ainsworth13,schiffer13} or small phase angle differences \cite{simpson13,schiffer14,schiffer14_2}. In addition, ideal clocks are usually assumed, {\em i.e.}, $\Gamma=\mathbf I_{|\mathcal N_1|}.$
Furthermore, whenever constant impedance or constant current loads are assumed, the algebraic equations in \eqref{mgo}, \eqref{powFlow} can be eliminated by an appropriate network reduction. This process is commonly known as Kron reduction and frequently employed in microgrid and power system studies. For further details on Kron reduction, the reader is referred to \cite{kundur94,doerfler13_3}.

\section{Conclusions and topics of future research}
\lab{sec7}
\subsection{Summary}
The present survey paper has introduced the reader to the microgrid concept with the main focus of providing a detailed procedure for the model derivation of a three-phase inverter-based microgrid. In particular, it has been shown  how---and under which assumptions---the microgrid models usually used in the literature can be obtained from a significantly more complex model derived from fundamental physical laws. The assumptions invoked in the reduction process are often satisfied in standard applications. Therefore, the reduced model represents a valid approximation and may, hence, be useful for control design and system analysis. In addition, the employed model reduction techniques can equivalently be applied to standard bulk power system models.

Nevertheless, it is important to note that the model derived in the present paper neglects effects such as asymmetric operation, DC-side dynamics of DG units or line capacitances. These facts have to be kept in mind, when performing microgrid analysis based on the derived model and assessing the results.

Also, it is worth mentioning that numerical simulation of the introduced microgrid models \eqref{overall}, \eqref{overall2}, respectively \eqref{mgo}, \eqref{powFlow}, requires careful selection of the numerical integration method to be employed. The main reason for this is that the model \eqref{overall}, \eqref{overall2} contains dynamics evolving at a wide range of time-scales, {\em i.e.}, it is a stiff model \citep[Chapter 7]{hoffman01}, \citep[Chapter 8]{fatunla14}. As is well-known, certain (standard) numerical integration methods will lead to {\em numerical} instability, when applied to stiff models---unless an extremely small step size is employed \citep[Chapter 7]{hoffman01}, \citep[Chapter 8]{fatunla14}. On the contrary, in the reduced model \eqref{mgo}, \eqref{powFlow} the fast dynamics have been eliminated and replaced by their corresponding steady-state equations. Hence, this model is not stiff and simpler integration methods can be used. Very similar situations are usually encountered in simulation of large conventional power systems \citep[Chapter 13]{machowski08}, \cite{demiray08}.

\subsection{Future research}
To conclude this survey, we very briefly highlight some topics of future research.
Following up on the discussion in the introduction, there are numerous challenges related to system and control theory in microgrid applications. To further motivate these, we briefly review control goals in microgrids. 
At the present, the following are considered to be among the most relevant control objectives in microgrids \cite{path_microgrid, lasseter02,green07,hatziargyriou07,katiraei08,glover11,guerrero13}: frequency stability, voltage stability, operational compatibility of inverter-interfaced and SG-interfaced units, desired power sharing in steady-state, seamless switching from grid-connected to islanded-mode and vice-versa, robustness with respect to uncertainties and optimal dispatch.

Some of these problems have been addressed in recent work within the control community, 
{\em e.g.}, frequency and voltage stability \cite{simpson12_2,simpson13,schiffer13_2,torres13,dhople14,ainsworth13,muenz14,schiffer14,schiffer14_2,doerfler14}, secondary control \cite{bidram13,simpson12_2,bidram14,doerfler14,shafiee14} or optimal dispatch \cite{doerfler14,bolognani13,hans14}. 
Compared to the model derived in the present paper, most of the aforementioned work is conducted under certain additional simplifying assumptions, such as constant voltage amplitudes, identical DG unit dynamics, as well as lossless or identical lines. 

On the modeling side, one direct extension of the presented modeling framework is to investigate the suitability of the use of more refined modeling techniques such as dynamic phasors \cite{stankovic00,maksimovic01,demiray08} or symmetric components \cite{paap00,glover11}, to describe the dynamics of a microgrid in asymmetric operating conditions. In this context, one main challenge is to accurately consider these phenomena while maintaining an analytically trac\-table model.
Another subject of future research is the derivation of more detailed load models for microgrid applications. 
As in conventional power systems, accurate load modeling is a very important, but also very difficult task \cite{kundur94,vancutsem98}. 
The main reason for this is that there are typically many different kinds of loads connected within one power system or microgrid, see, e.g., \citep[Chapter 7]{kundur94}.  
As a consequence, it is difficult to obtain suitable generically valid abstractions.

In conclusion, there are many challenging open research questions regarding a reliable, safe and efficient operation of microgrids. Therefore, the authors hope that the present survey on modeling of microgrids may serve as a stimulating base for a large variety of future research on both the theoretical and the application side.

\bibliographystyle{model5-names} 
\bibliography{bib_microgrids}

\end{document}